\documentstyle[prd,eqsecnum,graphics,aps]{revtex}

\newcommand{\half}{{\case{1}{2}}}
\newcommand{\om}{\omega}
\newcommand{\bom}{{\bar\omega}}
\newcommand{\ep}{\epsilon}
\newcommand{\tmr}{\frac{2M_0}{R}}
\newcommand{\rx}{\frac{r}{R}}

\newcommand{\sinY}{\sin\theta \partial_{\theta} Y_l^m}
\newcommand{\cosY}{\cos\theta Y_l^m}

\newcommand{\bi}{\begin{itemize}}
\newcommand{\ei}{\end{itemize}}
\newcommand{\be}{\begin{equation}}
\newcommand{\ee}{\end{equation}}
\newcommand{\beqa}{\begin{eqnarray}}
\newcommand{\eeqa}{\end{eqnarray}}
\newcommand{\ba}{\begin{array}}
\newcommand{\ea}{\end{array}}
\newcommand{\bea}{\begin{eqnarray}}
\newcommand{\eea}{\end{eqnarray}}
\newcommand{\bean}{\begin{eqnarray*}}
\newcommand{\eean}{\end{eqnarray*}}

\newcommand{\ds}{\displaystyle}

\newcommand{\nn}{\nonumber}
\newtheorem{thm}{Theorem}

\newcommand{\apjl}{Astrophys. J. Lett.}
\newcommand{\aap}{Astron. and Astrophys.}

\newcommand{\mnras}{ Mon. Not. Roy. Astr. Soc.}

\begin{document}
\preprint{CGPG-99/12-2}
\draft


\title{The rotational modes of relativistic stars, I: Analytic results}
\author{Keith H. Lockitch\thanks{Email address: lockitch@gravity.phys.psu.edu}}
\address{Center for Gravitational Physics and Geometry, Department of 
Physics, \\ Pennsylvania State University, 104 Davey Laboratory,
State College, PA 16802, USA}

\author{Nils Andersson\thanks{Email address: na@maths.soton.ac.uk}}
\address{Department of Mathematics, University of Southampton, \\ 
Southampton SO17 1BJ, United Kingdom}

\author{John L. Friedman\thanks{Email address: friedman@uwm.edu}}
\address{Department of Physics, 
University of Wisconsin-Milwaukee, \\ P.O. Box 413, Milwaukee, WI 53201, 
USA}

\date{\today}
\maketitle

\begin{abstract}
We study the r-modes and rotational ``hybrid'' modes of relativistic
stars. As in Newtonian gravity, the spectrum of low-frequency rotational
modes is highly sensitive to the stellar equation of state. If the star 
and its perturbations obey the same one-parameter equation of state (as 
with isentropic stars), there exist {\it no pure r-modes at all} - no modes
whose limit, for a star with zero angular velocity, is an axial-parity
perturbation.  Rotating stars of this kind similarly have no pure 
g-modes, no modes whose spherical limit is a perturbation with polar 
parity and vanishing perturbed pressure and density. 

In spherical stars of this kind, the r-modes and g-modes form a 
degenerate zero-frequency subspace.  We find that rotation splits the 
degeneracy to {\it zeroth} order in the star's angular velocity $\Omega$, 
and the resulting modes are generically hybrids, whose limit as 
$\Omega\rightarrow 0$ is a stationary current with both axial and polar parts. 
Because each mode has definite parity, its axial and polar
parts have alternating values of $l$.   We show that each 
mode belongs to one of two classes, axial-led or polar-led, 
depending on whether the spherical harmonic with the lowest 
value of $l$ that contributes to its velocity field is axial 
or polar. Newtonian isentropic stars retain a vestigial set of purely
axial modes (those with $l=m$); however, for relativistic isentropic 
stars we show that these modes must also be replaced by axial-led hybrids. 
We compute the post-Newtonian corrections to the $l=m$ modes for uniform 
density stars.

On the other hand, if the star is non-isentropic (or, more broadly, if
the perturbed star obeys an equation of state that differs from that of
the unperturbed star) the r-modes alone span the degenerate zero-frequency 
subspace of the spherical star. In Newtonian stars, this degeneracy is 
split only by the order $\Omega^2$ rotational corrections. However, when 
relativistic effects are included the degeneracy is again broken at zeroth 
order.  We compute the r-modes of a non-isentropic, uniform density model to 
first post-Newtonian order.
\end{abstract}

\pacs{}


\section{Introduction}

The discovery that the r-modes in rotating stars are 
generically unstable due to the emission of gravitational
waves \cite{a97,jfs97}
has attracted a large amount of attention in the last two years.
The current models suggest that the r-mode instability 
may cause a newly born neutron star to spin down to a fraction 
of the Kepler frequency (that provides the limit of
dynamical stability) in the first few months of its existence
\cite{lom98,aks98}. 
Since a considerable amount of gravitational radiation is generated in the 
process, the r-modes provide a promising source for the 
generation of gravitational wave interferometers that are currently
under construction \cite{o98}. It is also speculated that
the instability associated 
with the r-modes may be relevant for older neutron stars
in accreting systems \cite{bi98,akst98}. 

Since the instability was first discovered, and its potential
astrophysical relevance was appreciated, there have been many 
attempts to improve on the detailed physics incorporated
in the models. This effort leads to difficult
questions regarding, for example, neutron star superfluidity \cite{lm99},
the interplay between the magnetic field and fluid
pulsations \cite{s99,re99,hl99}, and the formation of a solid crust as 
a young neutron star cools \cite{bu00,akst00,rieu00,lu00,yl00b,wma00,lou00}. 
These, and several other, issues must
be addressed before the true astrophysical relevance
of the r-modes can be assessed. Our understanding of the r-mode
instability, however, is based almost entirely on
Newtonian calculations, and it is important to compute these modes 
in a relativistic context, where instability growth times may differ
significantly from Newtonian-based estimates.  (The closely
related instability of f-modes of rapidly
rotating stars is sharply strengthened by relativistic effects;
see \cite{ste98} for a review.)
\footnote{There are as yet no fully relativistic
calculations of other pulsation modes (like the f-mode)
of rapidly rotating stars, except in the Cowling approximation 
\cite{ye97}.}

The purpose of the present investigation is 
to understand how general relativity
affects the properties of the  r-modes.  
In order to address this issue we first need to discuss the 
general nature of modes of rotating stars. 

The spherical symmetry of a non-rotating star implies that its
perturbations can be divided into two classes, polar or axial, 
according to their behaviour under parity.  Where polar tensor 
fields on a 2-sphere can be constructed from the scalars $Y_l^m$
and their gradients $\nabla Y_l^m$ (and the metric on a 2-sphere),
axial fields involve the pseudo-vector $\hat r\times \nabla Y_l^m$, and
their behavior under parity is opposite to that of $Y_l^m$.  That is,
axial perturbations of odd $l$ are invariant under parity, and axial
perturbations with even $l$ change sign. Because a rotating star 
is also invariant under parity, its perturbations may also be divided
into distinct parity eigenstates.  If a mode varies continuously
along a sequence of equilibrium configurations that starts with a
spherical star and continues along a path of increasing rotation, the
mode will be called axial if it is axial for the spherical star.  Its
parity cannot change along the sequence, but $l$ is well-defined only
for modes of the spherical configuration.

It is useful to subdivide stellar pulsation modes according to the 
physics dominating their behaviour.  This classification was first 
developed by Cowling \cite{c41} for the polar perturbations of Newtonian 
polytropic models.  The p-modes of spherical models are polar-parity 
modes having pressure as their dominant restoring 
force\footnote{The lowest p-mode for each value of $l$ and $m$ is 
termed an f-mode or fundamental mode.}. They typically have large 
pressure and density perturbations and high frequencies (higher than 
a few kilohertz for neutron stars).  The g-modes are polar-parity modes 
that are chiefly restored by gravity.  They typically have very small 
pressure and density perturbations and low frequencies. Indeed, for 
spherical isentropic stars, which are marginally stable to convection, 
the g-modes are all zero-frequency and have vanishing perturbed pressure 
and density. Similarly, all axial-parity perturbations of non-rotating
perfect fluid 
models have zero frequency. The perturbed pressure and density as well 
as the radial component of the fluid velocity all vanish for axial 
perturbations; being rotational scalars, they must have polar parity.  
Thus, the axial 
perturbations of a spherical star are simply stationary horizontal fluid 
currents. This Newtonian picture of stellar pulsation
is readily generalised
to the relativistic case. The only difference is that the various modes
will now generate gravitational waves. This means that they
are no longer ``normal modes'', but satisfy
outgoing wave boundary conditions at spatial infinity.
Furthermore, one can identify an additional class of such 
outgoing modes in relativistic stars. Like the modes of black 
holes, these modes are essentially associated with the dynamical spacetime 
geometry and have been termed w-modes, or gravitational-wave modes 
\cite{ks86}. For a general discussion of the oscillations of 
relativistic stars we refer the reader to the recent review article by 
Kokkotas and Schmidt \cite{ks99}.

In general, the classification of modes is relevant also for 
rotating stars, even though the character of the various modes 
may be significantly affected by rotation. In particular, rotation
imparts a finite frequency to the zero-frequency perturbations of 
spherical stars. Because these modes are restored by the Coriolis 
force, their frequencies are proportional to the star's angular 
velocity, $\Omega$. In fluid mechanics, 
such modes are generally known as inertial modes \cite{g64,yl00,rieu00}. 
In non-isentropic stars these rotationally 
restored modes all have axial parity (the polar g-modes are nondegenerate
already in a spherical non-isentropic star because of internal
entropy gradients).  In astrophysics, these modes were first studied in 
Newtonian gravity by Papaloizou and Pringle \cite{pp78}, who called 
them r-modes because 
of their similarity to the Rossby waves of terrestrial meteorology. 
In isentropic stars, however, the space of zero frequency modes of 
the spherical model includes the polar g-modes in addition to the 
axial r-modes.  This large degenerate subspace of zero-frequency 
modes is split by rotation to zeroth order in the angular velocity, 
and the rotationally restored (inertial) modes of isentropic stars are 
generically hybrids whose spherical limits are mixtures of axial and 
polar perturbations.  This has been shown in Newtonian gravity by
Lockitch and Friedman \cite{lf99} (see also \cite{li98,lsvh92}).
In order to distinguish between the two classes of inertial modes
we refer to modes which become purely axial in the spherical 
limit as r-modes, while modes that limit to a mixed parity state
are called rotational hybrid modes. This is a natural nomenclature
given the standard distinction between axial and polar modes
in relativistic studies of spherical stars.  

Attempts to study the 
r-modes of rotating relativistic stars were not made
until rather recently \cite{a97,k98,bk99,kh99,lo99,kh00}.
In fact, the present investigation is the first 
study of this problem that puts all its different facets
in the proper context. In particular, we prove that
(apart from a set of stationary dipole modes) rotating 
relativistic isentropic stars have {\it no} pure 
r-modes (modes whose limit for a spherical star is purely axial).
This is in contrast with the 
isentropic Newtonian stars which retain a vestigial set of purely 
axial modes (those having spherical harmonic indices $l=m$). 
Instead, the Newtonian r-modes with $l=m\geq 2$ acquire
relativistic corrections with both axial and polar parity to become
discrete hybrid modes of the corresponding relativistic models. We 
compute these corrections for slowly rotating isentropic stars to first 
post-Newtonian order. 

For non-isentropic relativistic stars the situation is somewhat different.
In the slow-motion approximation in which they have so far been
studied,  non-isentropic stars have, remarkably, a {\it
continuous} spectrum. Kojima \cite{k98} has shown 
that purely axial modes would be described
by a single, second-order ODE for the modes' radial
behavior. He then argues that the continuous spectrum is implied by the
fact that eigenvalue problem is singular 
(the  coefficient of the highest derivative term of the
equation vanishes at some value of the radial coordinate).
This claim has been made mathematically precise by Beyer and 
Kokkotas \cite{bk99}.  
As the latter authors point out, the continuous 
spectrum may be an artifact of the vanishing of the imaginary 
part of the frequency in the slow rotation limit. (Or, more broadly, 
it may be an artifact of the slow rotation approximation itself.)  
In this paper we show that, in addition to the continuous spectrum, 
certain discrete modes also exist as solutions to Kojima's equation.
These modes are the relativistic analogue to the Newtonian
r-modes in non-isentropic stars. 
We compute these modes for slowly rotating non-isentropic stars.

In a complementary study of the relativistic r-modes, Kojima and 
Hosonuma \cite{kh00} have recently derived the order $\Omega^2$ 
rotational corrections to Kojima's equation. Working in the time 
domain, they derive a set of evolution equations for an axial 
perturbation and its lowest order polar and axial corrections. 
Direct numerical evolution of these equations 
(with appropriate initial data) would provide a useful 
comparison with our results on the modes of non-isentropic 
relativistic stars. 

When does one need to take into account the departure of a neutron star
from isentropy in computing rotational modes?   Because of bulk
viscosity, a gravitational-wave driven instability is unlikely to set
in above about $10^{10} K$.  This is well below the Fermi temperature
of the star's baryons, and the departure from isentropy appears to be
dominated by composition gradients in the crust and interior.  
These have been discussed in the context of
g-modes of spherical stars by Finn \cite{fi87} and by Reisenegger and
Goldreich \cite{rg92,rg94} and for rotating stars by Lai \cite{lai98}.  
Because the timescale of perturbations is too
slow to allow the beta- and inverse beta-decays that would allow a
displaced fluid element to adjust its composition to that of the
surrounding star, $\Delta p/\Delta \rho$ is greater than $d \log p/d
\log \rho$ by a factor $1+\frac12 x$, where $x = n_p/n \approx 6\times
10^{-3}\rho/\rho_{\rm nuclear}$  is the local ratio of protons to
baryons.  This leads in the star's interior to g-mode frequencies
limited by the Brunt-V\"ais\"al\"a-frequency, 
\[
g\left(\frac{3\rho}{10P} x\right)^{1/2} \sim 500 {\rm s}^{-1} 
(\frac\rho{\rho_{\rm nuclear}})^{1/2}; 
\] 
 when a crust is present, crustal g-modes have comparable frequencies.
The g-mode frequencies of spherical stars are then of order 100-200 Hz;
when this is smaller than the frequencies of the rotationally restored
modes of the isentropic models, one expects the isentropic
approximation to be valid.

The plan of the paper is as follows. We begin, in Section \ref{sect2}, 
with a brief 
review of the Eulerian and Lagrangian perturbation formalisms, both of 
which are used in the paper.  In Section \ref{sect3}, we consider the 
time-independent
perturbations of spherical relativistic stars and prove that the subspace 
of nonradial zero-frequency modes is spanned by the r- and g-modes in 
isentropic models, but by the r-modes alone in non-isentropic models.
Because of this difference, the character of the mode spectrum in rotating
isentropic models differs considerably from that of non-isentropic models. 
In Section \ref{sect4a}, we consider rotating non-isentropic stars and 
argue that the problem 
of finding their r-modes is well-defined. In Section \ref{sect4b}, we 
consider the
isentropic case and derive a set of perturbation equations whose structure 
parallels the corresponding Newtonian equations of Lockitch and Friedman
\cite{lf99}\footnote{We will refer to equations from Ref. \cite{lf99} by 
the equation number with the prefix ``LF''. For example, Eq. (LF,25) will 
mean equation (25) from Ref. \cite{lf99}.}.  This 
similarity between the Newtonian and relativistic equations leads to an 
identical structure of the mode spectrum and to a parallel theorem that 
every non-radial mode is either an axial-led or polar-led 
hybrid (the result has so far been proven only for slowly rotating 
relativistic stars). We consider the relativistic r-mode solutions of 
isentropic stars, finding that the zero-frequency dipole ($l=1$) solutions 
are the only purely axial solutions allowed. In other words, there are 
no nonstationary modes in isentropic relativistic stars whose limit as 
$\Omega\rightarrow 0$ is a pure axial perturbation. In particular, the 
Newtonian r-modes having $l=m\geq 2$ do not exist in isentropic 
relativistic stars and must be replaced by axial-led hybrid modes.  
This section concludes with a discussion of the boundary conditions 
appropriate to the relativistic modes (Section \ref{sect4c}).
Finally, in Section \ref{sect5}
we construct the post-Newtonian corrections to the well-known Newtonian 
r-modes in uniform density stars, both isentropic and non-isentropic.
Some of the detailed equations, as well as the 
proof of the theorem regarding the isentropic mode spectrum,
are presented in Appendices~A-C.  We use geometrized units
($G=c=1$) throughout the paper.

\section{Eulerian and Lagrangian Perturbations}
\label{sect2}

In general relativity, a complete description of a self-gravitating
perfect fluid configuration is provided by a spacetime with metric 
$g_{\alpha\beta}$, sourced by an energy-momentum tensor,
\be
T_{\alpha\beta} = (\epsilon+p)u_\alpha u_\beta + p g_{\alpha\beta},
\label{emom}\ee
where the fluid 4-velocity $u^\alpha$ is a unit timelike vector
field,
\be
u^\alpha u_\alpha = -1,
\label{unit_u}
\ee
and $\ep$ and $p$ are, respectively, the total energy density 
and pressure of the fluid as measured by an observer moving with 
4-velocity $u^\alpha$.  The metric and fluid variables satisfy an 
equation of state,
\be
\ep = \ep(p,s),
\ee
with $s$ the entropy per baryon, as well as the Einstein equation,
\be
G_{\alpha\beta}=8\pi T_{\alpha\beta}.
\label{einstein}
\ee

An equilibrium stellar model is a stationary solution
$(g_{\alpha\beta}, u^\alpha, \ep, p)$ to these equations. In this paper
we will
consider only equilibrium models obeying a barotropic (one-parameter)
equation of state,
\be
\ep = \ep(p),
\ee
because this accurately models the equilibrium configuration of a 
neutron star.

Adiabatic perturbations of such a star may be studied using 
either the Eulerian or the Lagrangian perturbation formalism 
\cite{f78,fi92}.  
An Eulerian perturbation may be described in terms of a smooth family, 
$[\bar g_{\alpha\beta}(\lambda), \bar u^\alpha(\lambda), 
\bar \ep(\lambda), \bar p(\lambda)]$, of
solutions to the exact equations (\ref{unit_u})-(\ref{einstein}) 
that coincides with the equilibrium solution at $\lambda=0$,
\[
[\bar g_{\alpha\beta}(0), \bar u^\alpha(0), \bar \ep(0), \bar p(0)] 
= (g_{\alpha\beta}, u^\alpha, \ep, p).
\]
Then the Eulerian change $\delta Q$ in a quantity $Q$ 
may be defined (to linear order in $\lambda$) as,
\be
\delta Q \equiv \left. \frac{dQ}{d\lambda}\right|_{\lambda=0}.
\ee
Thus, an Eulerian perturbation is simply a change 
$(h_{\alpha\beta}, \delta u^\alpha, \delta \ep, \delta p)$ 
in the equilibrium configuration at a particular point in 
spacetime (where we have written the change in the metric as 
$h_{\alpha\beta}\equiv\delta g_{\alpha\beta}$).  These must 
satisfy the perturbed Einstein equation
$\delta G_\alpha^{\ \beta} = 8\pi\delta T_\alpha^{\ \beta}$,
together with an equation of state relating $\delta \ep$ and
$\delta p$ that may, in general, differ from that of the 
equilibrium configuration (see Eq. (\ref{ad_osc}) below).

In the Lagrangian perturbation formalism \cite{f78,fi92}, 
on the other hand, perturbed quantities are expressed 
in terms of the Eulerian change in the metric, $h_{\alpha\beta}$ and 
a Lagrangian displacement vector $\xi^\alpha$, which connects fluid 
elements in the equilibrium star to the corresponding elements in the 
perturbed star.  The Lagrangian change $\Delta Q$ in a quantity 
$Q$ is related to its Eulerian change $\delta Q$ by
\be
\Delta Q = \delta Q + \pounds_{\xi} Q,
\label{DelQ}
\ee
with $\pounds_{\xi}$ the Lie derivative along $\xi^\alpha$. 

The identities,
\begin{eqnarray}
\Delta g_{\alpha\beta} &=& h_{\alpha\beta} 
+ 2\nabla_{\left(\alpha\right.}\xi_{\left.\beta\right)} 
\label{Del_metric} \\
\Delta\varepsilon_{\alpha\beta\gamma\delta} &=&
\half\varepsilon_{\alpha\beta\gamma\delta} \
g^{\mu\nu} \Delta g_{\mu\nu}
\end{eqnarray}
then allow one to express the fluid perturbation in terms of
$h_{\alpha\beta}$ and $\xi^\alpha$,
\be
\Delta u^\alpha = 
\half u^\alpha u^\beta u^\gamma\Delta g_{\beta\gamma}  
\ee
\be
\frac{\Delta p}{\Gamma_1 p} = \frac{\Delta \epsilon}{\epsilon + p}
= \frac{\Delta n}{n} = -\half q^{\alpha\beta}\Delta g_{\alpha\beta}
\label{Del_etc} \\
\ee
where $\Gamma_1$ is the adiabatic index, $n$ is the baryon density 
and $q^{\alpha\beta} \equiv g^{\alpha\beta} + u^\alpha u^\beta$.
Using Eqs. (\ref{DelQ})-(\ref{Del_etc}), it is 
straightforward to express the corresponding Eulerian changes
also in terms of $h_{\alpha\beta}$ and $\xi^\alpha$, e.g., 
\be
\delta u^\alpha = q^\alpha_{\ \beta}\pounds_u\xi^\beta 
+ \mbox{$\half$} u^\alpha u^\beta u^\gamma h_{\beta\gamma}.
\label{delu}
\ee
For an adiabatic perturbation of a barotropic equilibrium model,
Eqs. (\ref{DelQ}) and (\ref{Del_etc}) imply that the Eulerian changes 
in the pressure and energy density are related by
\be
\frac{\delta p}{\Gamma_1 p} = \frac{\delta \ep}{(\ep+p)} 
+ \xi^\alpha A_\alpha
\label{ad_osc}
\ee
where we have introduced the Schwarzschild discriminant,
\be
A_\alpha = \frac{1}{(\ep+p)}\nabla_\alpha\ep 
- \frac{1}{\Gamma_1 p}\nabla_\alpha p,
\ee
which governs convective stability in the star.
In general, the adiabatic index $\Gamma_1$ need not be the function
\be
\Gamma \equiv \frac{(\ep+p)}{p}\frac{dp}{d\ep}
\label{differ}
\ee
associated with the equilibrium equation of state. In terms of this
function we have
\be
A_\alpha = \left(\frac{1}{\Gamma} - \frac{1}{\Gamma_1}\right)
\frac{1}{p}\nabla_\alpha p .
\label{schwarz}
\ee
We will call a model isentropic if 
the perturbed configuration satisfies the same barotropic equation
of state as the unperturbed configuration. In this case,
$\Gamma_1\equiv\Gamma$ and the Schwarzschild discriminant vanishes 
identically. Such stars are marginally stable to convection.
In this paper we study low-frequency 
pulsation modes of slowly rotating relativistic stars. We
consider both isentropic and non-isentropic models.

\section{Stationary Perturbations of Spherical Stars}
\label{sect3}

The equilibrium of a spherical perfect fluid star is described
by a static, spherically symmetric spacetime with metric 
$g_{\alpha\beta}$ of the form,
\be
ds^2 = -e^{2\nu(r)} dt^2 + e^{2\lambda(r)} dr^2 + r^2 d \theta^2 
	+ r^2 sin^2 \theta d \varphi^2 \ ,
\ee
and the energy-momentum tensor (\ref{emom}) with the fluid
four-velocity given by
\be
u^\alpha = e^{- \nu} t^\alpha \ .
\ee
Here $t^\alpha=(\partial_t)^\alpha$  is
the timelike Killing vector of the spacetime.

For barotropic stars, 
the pressure and energy density are 
related by an equation of state of form
\be
p = p(\ep) \ .
\label{GR_sph:eos}
\ee
In addition to this, the various quantities must satisfy
the Einstein equation,
$G_{\alpha\beta}=8\pi T_{\alpha\beta}$, which leads to the
standard TOV equations,
\be
\frac{dp}{dr} = - \frac{(\ep+p)(M+4\pi r^3p)}{r(r-2M)} \ ,
\label{GR_sph:tov}
\ee
\be
\frac{dM}{dr} = 4\pi r^2\ep \ , 
\label{GR_sph:dMdr}
\ee
and
\be
\frac{d\nu}{dr} = - \frac{1}{(\ep+p)}\frac{dp}{dr} \ ,
\label{GR_sph:dnudr}
\ee
where
\be
M(r) \equiv \half r(1-e^{-2\lambda}) \ .
\label{GR_sph:mass_def}
\ee

Our main focus in this study is on the low-frequency 
oscillations, corresponding to 
 r-modes and hybrid modes, of slowly rotating stars.
As in Newtonian theory, we expect these modes to 
limit to stationary perturbations of a spherical star as the
rotation rate goes to zero. In other words, 
we are interested in the space of zero-frequency modes; 
the linearized, time-independent perturbations of the static
equilibrium.  As in the Newtonian case \cite{lf99}, we find that this 
zero-frequency subspace is spanned by two classes of perturbations.
To identify these classes explicitly, we must examine the 
equations governing the perturbed configuration.

Using the Eulerian formalism, we express the perturbed configuration 
in terms of the set
$(h_{\alpha\beta}, \delta u^\alpha, \delta \ep, \delta p)$,
satisfying the perturbed Einstein equation
$\delta G_\alpha^{\ \beta} = 8\pi\delta T_\alpha^{\ \beta}$,
together with an equation of state relating $\delta\ep$ and $\delta p$.

The perturbed Einstein tensor is given by
\beqa
\delta G_\alpha^{\ \beta}  &=& -\frac{1}{2} \biggl\{
  \nabla_\gamma \nabla^\gamma h_\alpha^{\ \beta}
- \nabla_\gamma \nabla^\beta h_\alpha^{\ \gamma} 
- \nabla^\gamma \nabla_\alpha h_\gamma^{\ \beta}
+ \nabla_\alpha \nabla^\beta h \nonumber \\
&+& 2 R_\alpha^{\ \gamma} h_\gamma^{\ \beta}
+  ( \nabla^\alpha \nabla^\beta h_{\alpha\beta}
  -\nabla_\gamma \nabla^\gamma h
  -R^{\alpha\beta} h_{\alpha\beta} )\delta_\alpha^{\ \beta}
\biggr\} 
\label{pert_Gmunu}
\eeqa
where $h\equiv g^{\alpha\beta}h_{\alpha\beta}$, $\nabla_\alpha$ is the 
covariant derivative associated with the equilibrium metric and
\be
R_\alpha^{\ \beta}
= 8\pi(T_\alpha^{\ \beta} - \half T \delta_\alpha^{\ \beta})
= 8\pi\left[(\ep+p)u_\alpha u^\beta 
+ \half (\ep-p)\delta_\alpha^{\ \beta}\right]
\ee
is the equilibrium Ricci tensor.
The perturbed energy-momentum tensor is given by
\beqa
\delta T_\alpha^{\ \beta} =
(\delta\ep+\delta p)u_\alpha u^\beta + \delta p \delta_\alpha^{\ \beta}
+ (\ep+p)\delta u_\alpha u^\beta+(\ep+p)u_\alpha \delta u^\beta.
\label{pert_Tmunu}
\eeqa

Following Thorne and Campolattaro \cite{tc67}, we expand our perturbed 
variables in scalar, 
vector and tensor spherical harmonics. The perturbed energy density 
and pressure are scalars and therefore must have polar parity
\be
\delta\ep = \delta\ep(r) Y_l^m,
\label{GR:del_ep}
\ee
\be
\delta p = \delta p(r) Y_l^m.
\label{GR:del_p}
\ee

The perturbed 4-velocity for a polar-parity mode can be written
\beqa
\delta u_P^{\alpha}=  \biggl\{
\half H_0(r) Y_l^m t^\alpha  
+ \frac{1}{r} W(r) Y_l^m r^{\alpha}  
+  V(r) \nabla^{\alpha} Y_l^m 
\biggr\} e^{-\nu}
\label{GR:del_u_po}
\eeqa
while that of an axial-parity mode can be written
\be
\delta u_A^{\alpha} = 
- U(r) e^{(\lambda-\nu)} \epsilon^{\alpha\beta\gamma\delta} 
\nabla_{\beta} Y_l^m u_\gamma \nabla_{\delta}\, r.
\label{GR:del_u_ax}
\ee
(We have chosen the exact form of these expressions for later convenience.)

To simplify the form of the metric perturbation we will again follow 
Thorne and Campolattaro \cite{tc67} and work in the 
Regge-Wheeler \cite{rw57} gauge. The metric perturbation 
for a polar-parity mode can then be written
\beqa
 h_{\mu\nu}^P = \left[
\ba[c]{cccc}
H_0(r) e^{2\nu} & H_1(r) & 0 & 0 \\
\mbox{\scriptsize symm} & H_2(r) e^{2\lambda} & 0 & 0 \\
0 & 0 & r^2 K(r) & 0\\
0 & 0 & 0 & r^2 sin^2\theta K(r)
\ea
\right] Y_l^m \ , 
\label{GR:h_po}
\eeqa
while that of an axial-parity mode can be written
\beqa
 h_{\mu\nu}^A =
 \left[
\ba[c]{cccc}
0 & 0 & h_0(r) \,(\frac{-1}{sin\theta})\partial_\varphi Y_l^m & h_0(r)\,\sinY\\
0 & 0 & h_1(r) \,(\frac{-1}{sin\theta})\partial_\varphi Y_l^m & h_1(r)\,\sinY\\
\mbox{\scriptsize symm} & \mbox{\scriptsize symm} & 0 & 0 \\
\mbox{\scriptsize symm} & \mbox{\scriptsize symm} & 0 & 0 \\
\ea
\right]
\label{GR:h_ax}
\eeqa

The Regge-Wheeler gauge is unique for perturbations having spherical
harmonic index $l\geq 2$.  However, when $l=1$ or $l=0$, there remain
additional gauge degrees of
freedom\footnote{Letting $e_{AB}$ be the metric on a two-sphere with
$\ep_{AB}$ and $D_A$ the associated volume element and covariant
derivative, respectively, one finds the following:  When $l\geq 2$ the 
polar tensors $D_{\! A}D_{\! B} Y_l^m$ and $e_{AB} Y_l^m$ are linearly 
independent, but when $l=1$, they coincide.  In addition, the axial tensor 
$\ep_{\left(A\right.}^{\ \ B}D_{\left.\! C\right)}D_{\! B} Y_l^m$ vanishes 
identically for $l=1$ and, of course, $D_{\! A} Y_l^m$ vanishes for $l=0$.}.  
In addition, the components of the perturbed Einstein equation acquire a 
slightly different form in each of the three cases. (Campolattaro and 
Thorne \cite{ct70} discuss the difference between the $l\geq 2$ and $l=1$ 
cases.)

We have derived the components of the perturbed Einstein equation
using the Maple tensor 
package\footnote{www.maplesoft.com}
by substituting expressions 
(\ref{GR:del_ep})-(\ref{GR:h_ax}) into Eqs.~(\ref{pert_Gmunu}) and 
(\ref{pert_Tmunu}) (making liberal use of the equilibrium equations
(\ref{GR_sph:tov}) through (\ref{GR_sph:mass_def}) to simplify the 
expressions).  
The resulting set of equations for the case $l\geq 2$ are equivalent 
to those presented by Thorne and Campolattaro \cite{tc67} upon 
specializing their equations to 
the case of stationary perturbations and making the necessary changes of 
notation\footnote{In particular, their fluid variables (denoted by the 
subscript ${\scriptstyle TC}$) are related to 
ours as follows: $U_{TC}(r,t)=Ut$, $W_{TC}(r,t)=-re^{\lambda}Wt$,
$V_{TC}(r,t)=Vt$ and $\delta\ep/(\ep+p)=\delta p/\Gamma_1 p=-(K+\half H_0)$.
 Their equilibrium metric has the opposite signature and differs 
in the definitions of the metric potentials
$\nu_{TC}=\half\nu$ and $\lambda_{TC}=\half\lambda$.}.  
Similarly, the set of equations for the case $l=1$ are equivalent to 
those presented by Campolattaro and Thorne \cite{ct70}.   
For completeness, the equations governing stationary 
perturbations of spherical stars are given in Appendix~A.

\subsection{Decomposition of the zero-frequency subspace.}

By inspection of the three sets of perturbation equations given in Appendix~A, 
it is evident that they decouple into two independent classes.
We find that any solution 
\be
(H_0, H_1, H_2, K, h_0, W, V, U, \delta\ep, \delta p),
\ee
to the equations governing the time-independent perturbations of 
a static, spherical star is a superposition of (i) a solution
\be
(0, H_1, 0, 0, h_0, W, V, U, 0, 0)
\ee
to Eqs. (\ref{tr_l2})-(\ref{tph_l2}) or (\ref{tr_l0}) and 
(ii) a solution
\be
(H_0, 0, H_2, K, 0, 0, 0, 0, \delta\ep, \delta p)
\ee
to Eqs. (\ref{tt_l2})-(\ref{rth_l2}), (\ref{tt_l1})-(\ref{rth_l1})
or (\ref{tt_l0})-(\ref{thth+_l0}).
 
For  solutions of type (ii), the vanishing of the $(tr)$, 
$(t\theta)$ and $(t\varphi)$ components of the perturbed metric 
in our coordinate system implies that these solutions are static.
If one assumes the linearization 
stability\footnote{We are aware of a proof of this linearization
stability property under assumptions on the equation of state that are 
satisfied by uniform density stars, but would not allow polytropes 
\cite{ks80}.}
of these solutions, i.e., that any solution to the static perturbation 
equations is tangent to a family of exact static solutions, then the 
theorem that any static self-gravitating perfect fluid is spherical 
implies that any solution of type (ii) is simply a neighboring spherical 
equilibrium. 

Thus, under the assumption of linearization stability we have shown that
all stationary non-radial ($l>0$) perturbations of a spherical star have
\[H_0=H_2=K=\delta\ep=\delta p=0\]
and satisfy Eqs. (\ref{tr_l2})-(\ref{tph_l2}). That is,
\bea
0 &=& H_1 +\frac{16\pi(\epsilon+p)}{l(l+1)} e^{2\lambda} r W,  
\label{GR:sph_H1} \\
&& \nn \\
0 &=& e^{-(\nu-\lambda)}\left[e^{(\nu-\lambda)}  H_1\right]'
+ 16\pi(\ep+p)e^{2\lambda}V,
\label{GR:sph_H1_2} 
\eea
\beqa
h_0^{''} - (\nu'+\lambda') h_0' 
+ \biggl[ \frac{(2-l^2-l)}{r^2}e^{2\lambda} 
- \frac{2}{r}(\nu'+\lambda')
- \frac{2}{r^2} \biggr] h_0 
= \frac{4}{r}(\nu'+\lambda') U
\label{GR:sph_h_0''}
\eeqa
where a prime denotes a derivative with respect to $r$.
Observe that if we use Eq. (\ref{GR:sph_H1}) to eliminate $H_1(r)$ from 
Eq. (\ref{GR:sph_H1_2}) we obtain
\be
V = \frac{e^{-(\nu+\lambda)}}{l(l+1)(\epsilon+p)} 
\left[(\epsilon+p)e^{\nu+\lambda} r W \right]'.
\label{GR:sph_V}
\ee
This equation is clearly the generalization to relativistic stars
of the conservation of mass equation in Newtonian gravity, Eq. (LF,13).
The other two equations relate the two dynamical degrees of freedom of 
the spacetime metric 
to the perturbation of the fluid 4-velocity and vanish in the Newtonian 
limit.

These perturbations must be regular everywhere and satisfy the boundary
condition that the Lagrangian change in the pressure vanishes at the 
surface of the star, $r=R$. We show in Section \ref{sect4c} below that 
this boundary condition requires only 
\be
W(R) = 0.
\ee
leaving $W(r)$ and $U(r)$ otherwise undetermined.  If $W(r)$ and 
$U(r)$ are specified, then the functions $H_1(r)$, $h_0(r)$ and 
$V(r)$ are determined by the above equations. The solutions for the
metric variables are
subject to matching conditions to the solutions in the exterior
spacetime, which must also be regular at infinity, see Section \ref{sect4c}.

Finally, we consider the equation of state of the perturbed star.
We have seen that for an adiabatic oscillation of a barotropic star 
Eq. (\ref{Del_etc}) implies that the perturbed pressure and energy 
density are related by
\be
\frac{\delta p}{\Gamma_1 p} = \frac{\delta \ep}{(\ep+p)} 
+ \xi^r \left[\frac{\ep'}{(\ep+p)} - \frac{p'}{\Gamma_1 p}
\right]
\label{ad_osc2}
\ee
for some adiabatic index $\Gamma_1(r)$.

The Lagrangian 
displacement vector $\xi^\alpha$ is related to our perturbation 
variables by Eq. (\ref{delu}),
\be
q^\alpha_{\ \beta} \pounds_u\xi^\beta = 
\delta u^\alpha - \mbox{$\ds{\half}$} u^\alpha u^\beta u^\gamma 
h_{\beta\gamma}
\ee
Thus, we have
\be
e^{-\nu}\partial_t\xi^r = \delta u^r
\ee
or [taking the initial displacement (at $t=0$) to be zero]
\be
\xi^r = t e^\nu \delta u^r.
\label{form_xi_r}
\ee

For the class of perturbations under consideration, we have
seen that $\delta p = \delta\ep = 0$. Thus Eqs. (\ref{ad_osc2}) and
(\ref{form_xi_r}) require that
\be
\delta u^r \left[\frac{\ep'}{(\ep+p)} 
- \frac{p'}{\Gamma_1 p}\right] = 0.
\label{eos_req}
\ee
For axial-parity perturbations this equation is automatically
satisfied, since $\delta u_A^\alpha$ has no $r$-component, cf. Eq.
(\ref{GR:del_u_ax}).  In other words,  a spherical barotropic star always 
admits a set of axial zero-frequency modes (the r-modes). 

For polar-parity perturbations, $\delta u^r_P = e^{-\nu} W(r)/r \neq 0$,
and Eq. (\ref{eos_req}) will be satisfied if and only if 
\be
\Gamma_1(r)\equiv\Gamma(r)=\frac{(\ep+p)}{p}\frac{dp}{d\ep}.
\ee
Thus, we see that a spherical barotropic star admits a 
class of zero-frequency modes (the g-modes)
if and only if the perturbed star obeys the same 
one-parameter equation of state as the equilibrium star.  Again, 
we call such stars isentropic, because isentropic models and 
their adiabatic perturbations obey the same one-parameter equation
of state. 

That all axial-parity fluid perturbations of a spherical
relativistic star are time-independent was first shown by Thorne and
Campolattaro \cite{tc67}.  
The time-independent g-modes in spherical, isentropic, relativistic 
stars were found by Thorne \cite{t69a}.

Summarizing our results, we have shown that a spherical
barotropic star always admits a class of zero-frequency r-modes 
(stationary fluid currents with axial parity); but admits
zero-frequency g-modes (stationary fluid currents with polar parity)
if and only if the star is isentropic.  Conversely, the zero-frequency
subspace of non-radial perturbations of a non-isentropic spherical 
star is spanned by the r-modes only; while the zero-frequency
subspace of non-radial perturbations of a spherical isentropic 
star is spanned by the r- and g-modes - that is, by convective fluid 
motions having both axial and polar parity and with vanishing perturbed 
pressure and density.  Being stationary, these r- and g-modes do not 
couple to gravitational radiation, although the r-modes do induce a 
nontrivial metric perturbation ($h_{t\theta}, h_{t\varphi}\neq 0$) 
in the spacetime exterior to the
star (frame-dragging).  One would expect this large subspace of 
modes, which is degenerate at zero-frequency, to be split by 
rotation, as it is in Newtonian stars. 
Our aim is to investigate this issue in more detail, and we will begin by
considering perturbations of slowly rotating relativistic stars.


\section{Perturbations of Slowly Rotating Stars}
\label{sect4}

The equilibrium of a perfect fluid star that is
rotating slowly with uniform angular velocity $\Omega$ is
described \cite{h67,cm74} by a stationary, axisymmetric 
spacetime with metric, $g_{\alpha\beta}$, of the form 
\beqa
ds^2 = -e^{2\nu(r)} dt^2 + e^{2\lambda(r)} dr^2 + r^2 d \theta^2 
+ 
r^2 sin^2 \theta d \varphi^2  - 2 \omega(r) r^2 sin^2\theta dt d\varphi
\label{equil_metric}
\eeqa
(accurate to order $\Omega$).  The energy-momentum tensor follows from
(\ref{emom}) and  the fluid 4-velocity to order $\Omega$;
\be
u^\alpha = e^{-\nu} ( t^\alpha + \Omega \varphi^\alpha)
\label{equil_4v}
\ee
Here  
$t^\alpha=(\partial_t)^\alpha$ and
$\varphi^\alpha=(\partial_\varphi)^\alpha$ are, respectively, the
timelike and rotational Killing vectors of the spacetime.

That the star is rotating slowly corresponds to the assumption that $\Omega$
is small compared to the Kepler velocity, 
$\Omega_K\sim\sqrt{M/R^3}$, the 
angular velocity at which the star is dynamically unstable to mass shedding
at its equator.  In particular, we neglect all quantities of order 
$\Omega^2$ or higher. To  order $\Omega$ the star retains
its spherical shape, because the centrifugal deformation of its figure
is an order $\Omega^2$ effect \cite{h67}.
This means that the equations 
(\ref{GR_sph:eos})-(\ref{GR_sph:mass_def}) governing a spherical star
remain relevant also for a slowly rotating equilibrium configuration.
In addition 
we need to solve an equation \cite{h67}
that determines the new metric function $\om(r)$ in terms of the
spherical metric functions $\nu(r)$ and $\lambda(r)$,
\be
\frac{e^{(\nu+\lambda)}}{r^4}\frac{d}{dr}
\left( r^4 e^{-(\nu+\lambda)} \frac{d\bom}{dr} \right)
- \frac{4}{r} 
\left(\frac{d\nu}{dr}+\frac{d\lambda}{dr}\right) \bom = 0
\label{hartle}
\ee
where
\be
\bom(r)\equiv\Omega-\om(r).
\label{bom}
\ee
This new metric variable is a quantity of order
$\Omega$ that governs the dragging of inertial frames
induced by the rotation of the star \cite{h67}.
Apart from the frame-dragging effect, however, the 
spacetime is unchanged from the spherical configuration.
Outside the star, Eq. (\ref{hartle}) has the solution,
\be
\bom = \Omega-\frac{2J}{r^3}
\label{hartle_ext_soln}
\ee
where $J$ is the angular momentum of the spacetime.
This relation can  be used to provide boundary conditions for 
$\bar{\omega}$ (and its derivative) at the surface of the star in 
terms of $\Omega$ and $J$. 
Specifically, the solution to Eq. (\ref{hartle}) is normalized by 
requiring that
\be
\bar\omega(R) + \frac{1}{3}R\bar{\omega}'(R) = \Omega \ .
\label{om_cond}
\ee
where $R$ is the radius of the star.

Note that $\bar{\omega}_c $ and 
$\bar{\omega}$ satisfy the inequalities $0 < \bar{\omega}_c \le
\bar{\omega}
\le \Omega$ (where an index $c$ denotes the value at the centre of the
star).  This means that $0 \le \omega \le \Omega -  \bar{\omega}_c$, and 
that $\Omega$, $\omega$ and $\bar{\omega}$ are positive
for all values of $r$. Defining a rescaled variable 
$\tilde{\omega} = \bar{\omega}/\Omega$, we have
$\tilde{\omega}_c =\bar{\omega}_c/\Omega \le  \tilde{\omega} \le 1$. 
Then, to linear order in $\Omega$, a single integration of (\ref{hartle}) 
suffices to determine the 
frame dragging for all $\Omega$ and a specific stellar model 
(a given equation of state and, say, the central density).  

We now consider the nonradial perturbations of these slowly rotating 
equilibrium models to linear order in $\Omega$.
Since the equilibrium spacetime is stationary and axisymmetric, 
we may decompose our perturbations into modes of the 
form\footnote{We will always choose $m \geq 0$ since the 
complex conjugate of an $m<0$ mode with real frequency $\sigma$ is an 
$m>0$ mode with frequency $-\sigma$.  Note that $\sigma$ is the 
frequency measured by an inertial observer at infinity.} 
$e^{i(\sigma t + m \varphi)}$.  
The perturbation equations have been written down in the Eulerian 
formalism by Kojima \cite{k92}, but we will find it convenient
to work also in the Lagrangian formalism. We therefore begin by expanding 
the perturbed density and pressure, the displacement vector $\xi^\alpha$ and  
the metric perturbations $h_{\alpha\beta}$ in tensor spherical harmonics.

The Eulerian change in the density and pressure may be written as
\be
\delta\ep = \sum_{l=m}^\infty \, \delta\ep_l(r) \, Y_l^m \, e^{i\sigma t}
\label{GR:del_ep_exp}
\ee
and
\be
\delta p = \sum_{l=m}^\infty \, \delta p_l(r) \, Y_l^m \, e^{i\sigma t}
\label{GR:del_p_exp}
\ee
respectively.

The Lagrangian displacement vector can be written
\beqa
\xi^{\alpha} \equiv \frac{1}{i\kappa\Omega} \sum_{l=m}^\infty 
\biggl\{ 
\ds{\frac{1}{r} W_l(r) Y_l^m r^{\alpha} 
+ V_l(r) \nabla^{\alpha} Y_l^m }
- \ds{i U_l(r) P^\alpha_{\ \mu} \epsilon^{\mu\beta\gamma\delta} 
\nabla_{\beta} Y_l^m \nabla_{\!\gamma} \, t \nabla_{\!\delta} \, r}
\biggr\} e^{i\sigma t} \ ,
\label{xi_exp}
\eeqa
where we have defined,
\be
P^\alpha_{\ \mu} \equiv e^{(\nu+\lambda)} 
\left( \delta^\alpha_{\ \mu}
- t_\mu \nabla^\alpha t
\right)
\ee
and introduced the comoving frequency,
\be
\kappa\Omega \equiv \sigma+m\Omega.
\ee
The exact form of expression (\ref{xi_exp}) has been chosen for later 
convenience. In particular, we have chosen a gauge in which $\xi_t\equiv 0$.  
Note also the chosen relative phase between the terms in (\ref{xi_exp}) with 
polar parity (those with coefficients $W_l$ and $V_l$) and the terms 
with axial parity (those with coefficients $U_l$).

Working in the Regge-Wheeler gauge, we express our metric perturbation as
\beqa
h_{\mu\nu} = e^{i\sigma t}  \sum_{l=m}^\infty   
 \left[ \ba[c]{cccc}
H_{0,l}(r)e^{2\nu}Y_l^m & H_{1,l}(r)Y_l^m & 
h_{0,l}(r) \,(\frac{m}{sin\theta})Y_l^m & i h_{0,l}(r)\,\sinY\\
H_{1,l}(r)Y_l^m & H_{2,l}(r)e^{2\lambda}Y_l^m & 
h_{1,l}(r) \,(\frac{m}{sin\theta})Y_l^m & i h_{1,l}(r)\,\sinY\\
\mbox{\scriptsize symm} & \mbox{\scriptsize symm} &
r^2K_l(r)Y_l^m & 0\\
\mbox{\scriptsize symm} & \mbox{\scriptsize symm} &
0 & r^2sin^2\theta K_l(r)Y_l^m
\ea
\right] 
\label{h_components}
\eeqa
Again, note the choice of phase between the polar-parity components
(those with coefficients $H_{0,l}$, $H_{1,l}$, $H_{2,l}$ and $K_l$) 
and the axial-parity components (those with coefficients
$h_{0,l}$ and $h_{1,l}$).

The use of the Lagrangian formalism introduces additional gauge
freedom into the problem of stellar perturbations. This freedom is
associated with a class of trivial displacements that leave all physical
quantities invariant \cite{cq76,ss77}.  One eliminates this gauge freedom by
restricting attention to the ``canonical'' displacements --- those that 
conserve vorticity in constant entropy surfaces \cite{fs78a,f78}. This
conservation law, known as Ertel's theorem, is essentially the curl of
the perturbed Euler equation and in general relativity has the form \cite{f78},
\be
\Delta \left\{ {\pounds}_u \omega_{\alpha\beta} 
- \frac{2}{n^2} \nabla_{\left[\alpha\right.} n 
\nabla_{\!\left.\beta\right]} p \right\} = 0
\ee
or,
\be
{\pounds}_u \mbox{$\Delta$} \omega_{\alpha\beta} = 
\frac{2}{n^2} \left\{ \nabla_{\left[\alpha\right.} \Delta n 
\nabla_{\!\left.\beta\right]} p + \nabla_{\left[\alpha\right.} n 
\nabla_{\!\left.\beta\right]} \Delta p \right\},
\label{ertel}
\ee
where
\be
\omega_{\alpha\beta} \equiv 
2 \nabla_{\left[\alpha\right.} 
\Biggl(\frac{\epsilon+p}{n} u_{\left.\beta\right]}\Biggr)
\ee
is the relativistic vorticity.  For our slowly rotating equilibrium 
star, Eq. (\ref{ertel}) can be written using Eqs. (\ref{Del_etc}) 
and (\ref{schwarz}) as,
\be
i\kappa\Omega e^{-\nu} \Delta \omega_{\alpha\beta} = 
\frac{2}{n} A_r \nabla_{\left[\alpha\right.} r 
\nabla_{\!\left.\beta\right]} \Delta p ,
\label{Del_om_NI}
\ee
since $A_\alpha=A_r\nabla_\alpha r$, cf. (\ref{schwarz}). 
Note that the three spatial components of the perturbed vorticity 
are not independent, being related by the identity
\be
\nabla_{\left[\alpha\right.}\Delta\omega_{\left.\beta\gamma\right]} = 0.
\label{not_ind}
\ee

We seek those modes that in the limit $\Omega\rightarrow 0$ belong to 
the zero-frequency subspace considered in the previous section. We have 
shown that such modes must have axial parity in non-isentropic stars, but 
may be either polar or axial in the isentropic case.  We will, therefore, 
require that our perturbation variables obey an ordering in powers of 
$\Omega$ that reflects this spherical limit,
\begin{displaymath}
U_l, h_{0,l}  \sim  \phantom{1} O(1)
\end{displaymath}
\begin{displaymath}
W_l, V_l, H_{1,l}  \sim   
\Biggl\{  \ba{ll}
O(1) & \mbox{isentropic stars} \\
O(\Omega^2) & \mbox{non-isentropic stars}
\ea 
\end{displaymath}
\be
H_{0,l}, H_{2,l}, K_l, h_{1,l}, \delta \ep_l, \delta p_l, \sigma  
\sim  \phantom{1} O(\Omega) \ .
\label{ordering}
\ee
In addition to the new equations that arise at order $\Omega$,
the zeroth order quantities must obey the zeroth order perturbation 
equations (\ref{GR:sph_H1}), (\ref{GR:sph_h_0''}) and (\ref{GR:sph_V}), 
for all $l$.  The degeneracy of the zero-frequency modes will be split 
at {\it zeroth} order if there is a subset of the $O(\Omega)$ equations 
that involves only the $O(1)$ variables.  While this occurs in 
Newtonian gravity only for isentropic stars \cite{lf99}, in general 
relativity it occurs also for non-isentropic stars.  

\subsection{The non-isentropic case}
\label{sect4a}

In a search for the relativistic r-modes, Kojima \cite{k98} has recently 
applied his general perturbation equations \cite{k92} to the case of
a mode whose spherical limit is purely axial. Accordingly, he assumes 
an ordering of his perturbation variables in powers of $\Omega$ that 
agrees with our non-isentropic ordering (although he does not distinguish 
between the isentropic and non-isentropic cases). Kojima then 
finds that the zeroth order equation, 
Eq. (\ref{GR:sph_h_0''}), is joined at order 
$\Omega$ by an additional pair of equations, which can be written,

\beqa
 l(l+1) \biggl\{ i(\sigma + m\omega) e^{-2\nu}\left[  h_{0,l}^\prime - 2  
{h_{0,l} \over r} \right] +
{ (l-1)(l+2) h_{1,l} \over
r^2} \biggr\}
- 2im\omega^\prime e^{-2\nu} h_{0,l} = 0 \ ,
\label{ax1}
\eeqa
and
\begin{eqnarray}
 && l(l+1) \biggl\{ i(\sigma+m\omega)e^{-2\nu} h_{0,l}-
 e^{-2\lambda} h_{1,l}^\prime
-  \left[ {2M\over r^2 } + 4\pi(p-\epsilon)r \right] 
h_{1,l} \biggr\} \nonumber \\ 
 &+& im \left[ 16\pi(p+\epsilon)r^2 \bar{\omega} e^{-2\nu} h_{0,l} 
-
2r\omega^\prime e^{-2\nu-2\lambda} h_{0,l} 
\right.
 + \left.
\omega^\prime r^2 e^{-2\nu-2\lambda}
h_{0,l}^\prime \right] \nonumber \\
 &-& 16\pi m\bar{\omega}(p+\epsilon)r^2 e^{-2\nu} U_l = 0 \ .
\label{ax2}
\end{eqnarray}

These two equations can be combined to give a second relation between 
the zeroth order variables $h_{0,l}$ and $U_l$,
\begin{equation}
\left[\sigma+m\Omega - {2m\bar{\omega} \over l(l+1)} \right] U_l + 
(\sigma+m\Omega) h_{0,l} = 0 \ . 
\label{hUrel}
\end{equation}
Kojima derived this equation from the perturbed Einstein equation, but
as we will see in Section \ref{sect4b} the equation can be written down 
immediately 
in the Lagrangian formalism as one of the spatial components of Eq. 
(\ref{Del_om_NI}), $\Delta\om_{\theta\varphi}=0$ [cf. Eq. (\ref{om_th_ph})].  
Inspection of the Eulerian equations (as, for example, 
given by Kojima \cite{k92}) appears to suggest that there is no 
other equation in addition to Eqs. (\ref{GR:sph_h_0''}) and (\ref{hUrel}) 
that involves only the zeroth order axial variables $h_{0,l}$ and $U_l$.
However, after a closer study we find that for isentropic stars 
there is, in fact, a third such equation, implying that the 
system is overdetermined.  While the 
existence of this third equation is obscured by the Eulerian formalism, 
it arises naturally in the Lagrangian framework as the other independent
spatial component of Eq. (\ref{Del_om_NI}). In non-isentropic stars, this
equation couples the $O(1)$ variables occuring on the LHS to the 
$O(\Omega)$ variables (such as the perturbed pressure and density) 
appearing on the RHS.  However, for isentropic stars the RHS vanishes 
identically (since $A_r\equiv 0$) and the third equation relating only the 
$O(1)$ variables emerges [cf. Eq. (\ref{om_r_th}) or (\ref{om_ph_r})].  

Hence, for isentropic stars the assumption that the mode is purely 
axial as $\Omega\rightarrow 0$ leads to an overdetermined system of 
equations. 
The appropriate spherical limit is therefore
one that also includes the polar
variables $W_l$, $V_l$ and $H_{1,l}$, as in (\ref{ordering}). For
non-isentropic stars, on the other hand, the r-mode assumption appears 
to be consistent.  Combining Eq. (\ref{hUrel}) with Eq. 
(\ref{GR:sph_h_0''}) gives Kojima's ``master'' equation for $h_{0,l}$,
\begin{eqnarray}
\biggl[ \sigma + m\Omega - { 2m\bar{\omega} \over l(l+1)} \biggr] 
 \biggl\{  e^{\nu-\lambda} {d\over dr} \biggl[ e^{-\nu-\lambda} 
{dh_{0,l} \over dr} \biggr] 
&-& \biggl[{l(l+1) \over r^2} - {4M\over r^3}
+ 8\pi(p+\epsilon) \biggr]  h_{0,l} \biggr\} \nonumber \\
&& \nonumber \\
&+&
16\pi(p+\epsilon)(\sigma+m\Omega) h_{0,l} = 0 \ .
\label{singeq}
\end{eqnarray}

Kojima used this equation to argue that the r-mode spectrum of a 
relativistic star is continuous. The conclusion that the equation supports 
a continuous spectrum was shown with more mathematical rigour by 
Beyer and Kokkotas \cite{bk99}. Basically, the continuous spectrum arises 
because
(\ref{singeq}) is a singular eigenvalue problem; the combination
$ \sigma +m\Omega - { 2m\bar{\omega} / l(l+1)}$ may have a zero 
in the interval $r\in[0,\infty]$. It is interesting to ask whether
the presence of a continuous part of the spectrum is 
physical or whether it is an artifact of the approximations we have introduced.
That the latter may be the case can be argued for in the following way. 
To leading order in the slow-rotation expansion the mode frequency
$\sigma$ is a real valued quantity, but at higher orders it must
have also an imaginary part (corresponding to dissipation due to gravitational
wave emission). If we were to consider (\ref{singeq}) for complex frequencies, 
the problem will be regular and there will likely be no continuous spectrum.
The possible presence of a continuous spectrum 
is an interesting issue that should be investigated in more detail, but 
it is not the focus of the present study. 
What we want to emphasize here is that two important questions 
regarding equation (\ref{singeq}) have not yet been answered.
First of all, it has not been shown that the problem 
is well-defined. As we have already stated, one can show that the system of 
equations is overdetermined for isentropic stars. This means that, 
equation (\ref{singeq}) can only be relevant for non-isentropic stars.
But in order to show that the equation is, indeed, relevant we 
must demonstrate  
that all other perturbation variables are uniquely specified 
given a solution for $h_{0,l}$  from Eq. (\ref{singeq}). 
Given the relative complexity of the perturbed Einstein equation, 
this is not a trivial task.  
Secondly, we need to investigate whether Eq. (\ref{singeq}) 
allows distinct mode solutions in addition to its 
continuous spectrum.
After all, the true relativistic analogue to a Newtonian r-mode
ought to be a distinct mode with a well-defined eigenfunction. 

We address the first issue by considering the perturbation equations
that arise in the Eulerian formalism, cf. \cite{k92}.
As far as the axial perturbation variables are concerned, 
the  set of equations (\ref{ax1}), (\ref{hUrel}) and (\ref{singeq}) makes 
sense: 
We have three equations governing
$h_{0,l}$, $h_{1,l}$ and $U_l$, for all $l$. What is not so clear is 
whether the remaining $O(\Omega)$ perturbation equations yield unique 
$K_l$, $H_{0,l}$, $H_{2,l}$, $\delta p_l$ and $\delta \ep_l$ once the 
axial variables are specified. (For non-isentropic stars, the variables 
$W_l$, $V_l$ and $H_{1,l}$ are assumed to be of 
order $\Omega^2$ so they will not enter 
into our calculation). In order to show that this is the case we must 
show that the remaining equations can be reduced to five independent 
ones. In this effort we are immediately helped by the fact that, given the 
assumed ordering (\ref{ordering}); i) the two equations 
$\delta G_{t\varphi} = 8\pi \delta T_{t\varphi}$ and 
$\delta G_{t\theta} = 8\pi \delta T_{t\theta}$ 
both imply (\ref{GR:sph_h_0''}) at order $\Omega$, 
and ii) $\delta G_{tr} - 8\pi \delta T_{tr} \sim \Omega^2$ 
so  is automatically 
satisfied at lower orders.  
This leaves us with six equations: The equation 
of state for the perturbations and, for example, the five remaining Einstein 
equations. In other words, the problem would seem to be overdetermined. 
However, for non-isentropic stars the equation of state 
(\ref{ad_osc}) that relates 
$\delta p$ to $\delta \rho$ is of order $\Omega^2$. 
Thus, we have five equations for our five unknown
variables and the problem is well-defined. In other words, 
if a discrete mode which limits to a purely axial 
perturbation as $\Omega \to 0$ exists it should follow
from (\ref{singeq}).
For completeness, 
the perturbation equations for non-isentropic stars (complete
to order $\Omega$) that follow from (\ref{ordering}) are listed
in Appendix~B. 

Let us now suppose that a distinct r-mode solution exists in the  
non-isentropic case. One would intuitively expect this to be the case 
since there will then be an internal stratification in the star 
associated with the entropy gradient. In the Newtonian case, 
this stratification leads to a single r-mode for each 
combination of $l$ and $m$ at order $\Omega$ (these modes then become 
non-degenerate at order $\Omega^2$ \cite{pp78,pea81,s82}),
and it also leads to the presence of non-trivial 
polar g-modes. 

From the above discussion we know that a
relativistic r-mode of a non-isentropic star should follow 
from (\ref{singeq}). 
We begin our search for such 
solutions by deriving a constraint on the possible mode-frequencies. 
We do this by first
scaling out both $\Omega$ and $m$ from the problem by 
expressing the frequency $\sigma$ in terms of a 
real  parameter $\alpha$, such that
\begin{equation}
\sigma = -m\Omega\left[ 1 - {2\alpha\over l(l+1)} \right] \ .
\end{equation}
Then (\ref{singeq}) can be written 
\begin{equation}
(\alpha -\tilde{\omega})
\left\{  e^{\nu-\lambda} {d\over dr} \left[ e^{-\nu-\lambda} 
{dh_{0,l} \over dr} \right] - \left[{l(l+1) \over r^2} - {4M\over r^3}
+8\pi(p+\epsilon) \right]  h_{0,l} \right\}  +
16\pi(p+\rho) \alpha h_{0,l} = 0 \ ,
\label{singeq2}\end{equation}
where we have used $\tilde{\omega}= \bar{\omega}/\Omega$.
From this equation we see that
the eigenvalues $\alpha$ and the corresponding 
eigenfunctions $h_{0,l}$ are not explicitly dependent on either $\Omega$
or $m$. The latter means that, if we find an acceptable mode-solution
to (\ref{singeq2}) it will be relevant for all $m\neq 0$ for each given 
multipole $l$. This would be in accord with the non-isentropic
Newtonian case
where one finds a single r-mode for each combination of 
$l$ and $m$ at order $\Omega$ \cite{pp78,pea81,s82}.  

As we will now establish,
non-trivial solutions to (\ref{singeq2}) may exist provided that 
  $\alpha-\tilde{\omega}$  vanishes at
at least one point in the interval $r\in [0,\infty]$.
To show this we first assume that $\alpha-\tilde{\omega}$ does not
have a zero in $r\in [0,\infty]$.
Then we can define a new well-behaved function $f$ through
$h_{0,l} =r^2(\alpha-\tilde{\omega})f$. By introducing this definition in
(\ref{singeq2}),  multiplying
by $r^2 f$ and integrating over $r\in
[0,\infty]$
one can show that (as long as $f$ vanishes both as $r\to 0$ and 
$r\to \infty$ as is required by the  regularity conditions)
\begin{equation}
-\int_0^\infty (\alpha -\tilde{\omega})^2 r^4 e^{-\lambda-\nu}
| f^\prime | dr \nonumber = \int_0^\infty  (\alpha -\tilde{\omega})^2
[l(l+1)-2]r^2  e^{\lambda-\nu} |f|^2 dr \ .
\label{integ}\end{equation}
Here both integrands are positive definite, and it follows that we can have
no non-trivial solutions for $f$. 

In other words, a non-trivial
solution for $h_{0,l}$ can only exist
if  $\alpha-\tilde{\omega} = 0$ at  some point in the spacetime. 
That is, the eigenvalue
$\alpha$ must lie somewhere in the range 
\begin{equation}
\tilde{\omega}_c \le \alpha
\le 1 \ .
\label{range}\end{equation}
As already noticed by Kojima \cite{k98} this agrees well with the Newtonian 
result. As the star becomes less relativistic  $\tilde{\omega}_c\to 1$
and our integral relation then predicts a non-trivial solution
only for $\alpha =1$, i.e. the Newtonian r-mode eigenvalue.
We will attempt to find discrete r-mode solutions, with 
frequencies in the permissible range, in Section \ref{sect5b}.

\subsection{The isentropic case}
\label{sect4b}

As indicated above, the conservation of vorticity gives rise to a 
mixing of axial and polar modes at zeroth order in $\Omega$.
This suggests that the modes of isentropic stars will generically
be of a hybrid nature, and as a consequence the equations determining 
the modes are more complicated than those for r-modes of 
non-isentropic stars.  

The relevant perturbation equations for the isentropic case
follow from the spatial components of Eq. 
(\ref{Del_om_NI}), which for isentropic stars becomes simply,
\be
\Delta \omega_{\alpha\beta} = 0.
\label{Del_om}
\ee

We begin by expressing this relation, i.e.
\be
0 = \Delta \omega_{\alpha\beta} = 
\nabla_\alpha\left[\Delta\left(\frac{\epsilon+p}{n} u_\beta 
\right)\right]
- \nabla_\beta\left[\Delta\left(\frac{\epsilon+p}{n} u_\alpha 
\right)\right],
\ee
in terms of $\xi^\alpha$ and $h_{\alpha\beta}$.

Making use of Eq. (\ref{Del_etc}) we have
\be
\Delta \left( \frac{\epsilon+p}{n} u_\alpha \right) = 
\frac{\epsilon+p}{n} \left[ \Delta u_\alpha 
- \half q^{\alpha\beta}\Delta g_{\alpha\beta}\left( 
	\frac{\Gamma_1 p}{\epsilon+p}\right)u_\alpha 
\right],
\label{Del_hu}
\ee
where
\be
\Delta u_\alpha\equiv \Delta(g_{\alpha\beta}u^\beta)
=\Delta g_{\alpha\beta}u^\beta+g_{\alpha\beta}\Delta u^\beta
\ee

The ordering (\ref{ordering}) implies that 
$u^\alpha u^\beta h_{\alpha\beta}$ and 
$g^{\alpha\beta}h_{\alpha\beta}$ vanish to zeroth order 
in $\Omega$, since the only zeroth order metric components are
$h_{tr}$, $h_{t\theta}$ and $h_{t\varphi}$.
Therefore,
\begin{eqnarray}
\half u^\alpha u^\beta \Delta g_{\alpha\beta} 
	&=& u^\alpha u^\beta \nabla_\alpha \xi_\beta  \\
\half q^{\alpha\beta}\Delta g_{\alpha\beta} 
	&=& q^{\alpha\beta}\nabla_\alpha \xi_\beta
\label{Del_q}  \\
\Delta u^\alpha &=& 
u^\alpha u^\beta u^\gamma \nabla_\beta\xi_\gamma \\
\Delta u_\alpha &=& 
h_{\alpha\beta}u^\beta + 
	u^\beta\nabla_\beta\xi_\alpha 
	+ u^\beta\nabla_\alpha\xi_\beta 
	+ u_\alpha u^\beta u^\gamma\nabla_\beta\xi_\gamma.
\label{Del_u}
\end{eqnarray}

From Eqs. (\ref{Del_etc}) and (\ref{GR_sph:dnudr}) and the relation,
\bea
u^\alpha u^\beta \nabla_\alpha\xi_\beta 
&=& 
- \xi^\beta u^\alpha\nabla_\alpha u_\beta 
+ u^\alpha \nabla_\alpha(u^\beta \xi_\beta) \nn \\
&=& - \xi^\beta \nabla_\beta \nu + O(\Omega),
\eea
we obtain,
\begin{eqnarray}
\half q^{\alpha\beta}\Delta g_{\alpha\beta} 
	&=& \left(\frac{\epsilon+p}{\Gamma_1 p}\right)\nu' 
		e^{-2\lambda}\xi_r 
\label{form_Dp}  \\
u^\alpha u^\beta \nabla_\alpha\xi_\beta
	&=& - \nu' e^{-2\lambda}\xi_r
\end{eqnarray}
to zeroth order in $\Omega$.
We will also use the explicit form of $u_\varphi$ determined from 
Eq. (\ref{equil_4v}),
\be
u_\varphi = e^{-\nu}{\bar\omega}r^2\sin^2\theta,
\label{u_phi}
\ee
and the components of $\Delta u_\alpha$ to zeroth order in 
$\Omega$,
\begin{eqnarray}
\Delta u_r       &=& e^{-\nu} \biggl[
h_{tr}+i\kappa\Omega\xi_r
+\Omega r^2\partial_r\left(\frac{1}{r^2}\xi_\varphi\right)
+\frac{e^{2\nu}}{r^2}\partial_r\left(r^2\omega 
e^{-2\nu}\right)\xi_\varphi
\biggr] \\
\Delta u_\theta  &=& e^{-\nu} \biggl[
h_{t\theta}+i\kappa\Omega\xi_\theta
+\Omega\partial_\theta\xi_\varphi
-2{\bar\omega}\cot\theta\xi_\varphi
\biggr] \\
\Delta u_\varphi &=& e^{-\nu} \biggl[
\ba[t]{l}
h_{t\varphi}+i\kappa\Omega\xi_\varphi
+\Omega\partial_\varphi\xi_\varphi
+2{\bar\omega}\sin\theta\cos\theta\xi_\theta 
\vphantom{\frac{x^a_b}{y^a_b}}  \\
+e^\nu\partial_r\left(r^2{\bar\omega}e^{-\nu} 
\right)\sin^2\theta e^{-2\lambda}\xi_r
\vphantom{\frac{x^a_b}{y^a_b}}
\biggr]. 
\ea
\label{Del_u_cov_compts}
\end{eqnarray}
For completeness, we explicitly write down the components of 
$i\kappa\Omega{\vec\xi}$ to zeroth order in $\Omega$, cf. Eq. (\ref{xi_exp}),
\begin{eqnarray}
\label{xi_components}
&i\kappa\Omega\xi^t = O(\Omega)  &
i\kappa\Omega\xi^\theta = \ds{\sum_l \frac{1}{r^2\sin\theta} 
			\left[V_l \sinY + m U_l Y_l^m \right]e^{i\sigma t}} 
\nonumber \\
 && \nonumber \\
&i\kappa\Omega\xi^r = \ds{\sum_l \frac{1}{r} W_l Y_l^m e^{i\sigma t}} 
\phantom{12345} &
i\kappa\Omega\xi^\varphi = \ds{\sum_l \frac{i}{r^2\sin^2\theta}    
			\left[m V_l Y_l^m + U_l \sinY \right]e^{i\sigma t}}  
\nonumber \\
 &&  \nonumber \\
&i\kappa\Omega\xi_t = 0 &
i\kappa\Omega\xi_\theta = \ds{\sum_l \frac{1}{\sin\theta} 
			\left[V_l \sinY + m U_l Y_l^m \right]e^{i\sigma t}}  \nonumber \\
 &&  \nonumber \\
&i\kappa\Omega\xi_r = \ds{\sum_l \frac{e^{2\lambda}}{r} W_l Y_l^m e^{i\sigma t}} \phantom{12345} &
i\kappa\Omega\xi_\varphi = \ds{\sum_l i \left[m V_l Y_l^m 
+ U_l \sinY \right]e^{i\sigma t}}.
\end{eqnarray}


By making use of Eqs. (\ref{Del_hu}) through 
(\ref{Del_u_cov_compts}) and the expressions (\ref{xi_components}) 
and (\ref{h_components}) for the components of 
$i\kappa\Omega{\vec\xi}$ and $h_{\alpha\beta}$, we may now 
write the spatial components of $\Delta\omega_{\alpha\beta}$. 
We will use Eq. (\ref{GR:sph_H1}) to eliminate $H_{1,l}$ (for all $l$)
from the resulting expressions and drop the ``0'' subscript on the metric
functions $h_{0,l}$, writing $h_{0,l}\equiv h_l$.  


\begin{eqnarray}
\Delta\omega_{\theta\varphi}
 &=& \left(\frac{\epsilon+p}{n}\right) \biggl\{
\partial_\theta\Delta u_\varphi 
- \partial_\varphi\Delta u_\theta
- \partial_\theta\left[
\half q^{\alpha\beta}\Delta g_{\alpha\beta}\left( 
\frac{\Gamma_1 p}{\epsilon+p}\right)u_\varphi
\right]\biggr\} 
\nonumber \\
 &=& \left(\frac{\epsilon+p}{n}\right) 
\frac{e^{-\nu}\sin\theta}{i\kappa\Omega} 
\label{om_th_ph_form1}  \\
 & & \times\sum_l\biggl\{
	\ba[t]{l}
	\left[ l(l+1)\kappa\Omega(h_l+U_l)
	-2m{\bar\omega}U_l\right] Y_l^m \\
	\\
	-2{\bar\omega}V_l\left[\sinY+l(l+1)\cosY\right] \\
	\\
	+\frac{e^{2\nu}}{r}\partial_r\left(r^2{\bom}e^{-2\nu}
	\right)W_l\left[\sinY+2\cosY\right]
	\biggr\}e^{i\sigma t}
	\ea
\nonumber 
\end{eqnarray}


\begin{eqnarray}
\Delta\omega_{r\theta}
 &=& \left(\frac{\epsilon+p}{n}\right) e^\nu \biggl[
\partial_r\left(e^{-\nu}\Delta u_\theta\right) 
- \partial_\theta\left(e^{-\nu}\Delta u_r\right)
\biggr]
\nonumber\\
 &=& \left(\frac{\epsilon+p}{n}\right) 
\frac{e^\nu}{\kappa\Omega\sin\theta}
\label{om_r_th_form1}  \\
 & & \times\sum_l\Biggl\{
\ba[t]{l}
	m\kappa\Omega\partial_r\left[e^{-2\nu}(h_l+U_l)\right]Y_l^m 
	-2\partial_r\left({\bar\omega}e^{-2\nu}U_l\right)\cos\theta\sinY \\
	\\
	+\frac{1}{r^2}\partial_r\left(r^2{\bar\omega}e^{-2\nu}\right)U_l
	\left[m^2+l(l+1)(\cos^2\theta-1)\right]Y_l^m \\
	\\
	-2m\partial_r\left({\bar\omega}e^{-2\nu}V_l\right)\cos\theta Y_l^m
	+\frac{m}{r^2}\partial_r\left(r^2{\bar\omega}e^{-2\nu}\right)V_l\sinY \\
	\\
	+\kappa\Omega\left[\partial_r\left(e^{-2\nu}V_l\right)
	+e^{-2\nu}\left(\frac{16\pi r(\epsilon+p)}{l(l+1)}
	-\frac{1}{r}\right)e^{2\lambda}W_l
	\right]\sinY
	\Biggr\}e^{i\sigma t}
	\ea
\nonumber
\end{eqnarray}


\begin{eqnarray}
\Delta\omega_{\varphi r}
 &=& \left(\frac{\epsilon+p}{n}\right) e^\nu 
\biggl\{ \ba[t]{l}
	\partial_\varphi\left(e^{-\nu}\Delta u_r\right) 
	- \partial_r\left(e^{-\nu}\Delta u_\varphi\right) \\
	\\
	+ \partial_r\left[
	\half q^{\alpha\beta}\Delta g_{\alpha\beta}
	\left( \frac{\Gamma_1 p}{\epsilon+p}\right) e^{-\nu}u_\varphi
\right]\biggr\}
\ea
\nonumber \\
 &=& \left(\frac{\epsilon+p}{n}\right) \frac{e^\nu}{i\kappa\Omega}
\label{om_ph_r_form1} \\
 & & \times\sum_l\Biggl\{
\ba[t]{l}
	m\kappa\Omega\left[\partial_r\left(e^{-2\nu}V_l\right)
	+e^{-2\nu}\left(\frac{16\pi r(\epsilon+p)}{l(l+1)}
	-\frac{1}{r}\right)e^{2\lambda}W_l
	\right]Y_l^m \\
	\\
	-2\partial_r\left({\bar\omega}e^{-2\nu}V_l\right)\cos\theta\sinY
	+\frac{m^2}{r^2}\partial_r\left(r^2{\bar\omega}e^{-2\nu}\right)V_lY_l^m \\
	\\
	+\partial_r\left[\frac{1}{r}\partial_r\left(r^2{\bom}e^{-2\nu}\right)W_l
	\right](\cos^2\theta-1)Y_l^m \\
	\\
	-2m\partial_r\left({\bar\omega}e^{-2\nu}U_l\right)\cosY
	+\kappa\Omega\partial_r\left[e^{-2\nu}(h_l+U_l)\right]\sinY \\
	\\
	+\frac{m}{r^2}\partial_r\left(r^2{\bar\omega}e^{-2\nu}\right)U_l\sinY
	\Biggr\}e^{i\sigma t}
	\ea
\nonumber
\end{eqnarray}


These equations can be rewritten using the standard identities  
\begin{eqnarray}
\sin\theta\partial_\theta Y_l^m &=& l Q_{l+1} Y_{l+1}^m 
	- (l+1) Q_l Y_{l-1}^m \\
\cos\theta Y_l^m &=& Q_{l+1} Y_{l+1}^m + Q_l Y_{l-1}^m 
\end{eqnarray}
with $Q_l$ defined as,
\be
Q_l \equiv \left[ \frac{(l+m)(l-m)}{(2l-1)(2l+1)} \right]^{\half}. 
\label{Q_l}\ee

\noindent
We then get, 
from $\Delta\omega_{\theta\varphi}=0$,
\be
0 = \sum_l\Biggl\{
\ba[t]{l}
\left[ l(l+1)\kappa\Omega(h_l+U_l)
-2m{\bar\omega}U_l\right] Y_l^m \\
 \\
+\left[\frac{e^{2\nu}}{r}\partial_r\left(r^2{\bar\omega}e^{-2\nu}\right)W_l
-2l{\bar\omega}V_l\right](l+2)Q_{l+1}Y_{l+1}^m  \\
 \\
-\left[\frac{e^{2\nu}}{r}\partial_r\left(r^2{\bar\omega}e^{-2\nu}\right)W_l
+2(l+1){\bar\omega}V_l\right](l-1)Q_lY_{l-1}^m
\Biggr\}
\ea
\label{om_th_ph_form2}
\ee


\noindent
From $\Delta\omega_{r\theta}=0$ we have,
\be
0 = \sum_l\Biggl\{
\ba[t]{l}
\left[-2\partial_r\left({\bar\omega}e^{-2\nu}U_l\right)
+\frac{(l+1)}{r^2}\partial_r\left(r^2{\bar\omega}e^{-2\nu}\right)U_l
\right] l Q_{l+1}Q_{l+2}Y_{l+2}^m \\
 \\
+\biggl[
	\ba[t]{l}
	l\kappa\Omega\partial_r\left(e^{-2\nu}V_l\right)
	-2m\partial_r\left({\bar\omega}e^{-2\nu}V_l\right) \\
	 \\
	+\frac{lm}{r^2}\partial_r\left(r^2{\bar\omega}e^{-2\nu}\right)V_l
	+l\kappa\Omega e^{-2\nu}\left(\frac{16\pi r(\epsilon+p)}{l(l+1)}
					-\frac{1}{r}\right)e^{2\lambda}W_l
	\biggr]Q_{l+1}Y_{l+1}^m 
	\ea \\
 \\
+\biggl[
	\ba[t]{l}
	m\kappa\Omega\partial_r\left[e^{-2\nu}(h_l+U_l)\right]
	+2\partial_r\left({\bar\omega}e^{-2\nu}U_l\right)
	\left((l+1)Q_l^2-l Q_{l+1}^2\right) \\
	 \\
	+\frac{1}{r^2}\partial_r\left(r^2{\bar\omega}e^{-2\nu}\right)U_l
	\left[m^2+l(l+1)\left(Q_{l+1}^2+Q_l^2-1\right)\right]
	\biggr] Y_l^m 
	\ea \\
 \\
-\biggl[
	\ba[t]{l}
	(l+1)\kappa\Omega\partial_r\left(e^{-2\nu}V_l\right)
	+2m\partial_r\left({\bar\omega}e^{-2\nu}V_l\right) \\
	 \\
	+\frac{m(l+1)}{r^2}\partial_r\left(r^2{\bar\omega}e^{-2\nu}\right)V_l
	+(l+1)\kappa\Omega e^{-2\nu}\left(\frac{16\pi r(\epsilon+p)}{l(l+1)}
					-\frac{1}{r}\right)e^{2\lambda}W_l
	\biggr]Q_l Y_{l-1}^m 
	\ea \\
 \\
+\left[2\partial_r\left({\bar\omega}e^{-2\nu}U_l\right)
+\frac{l}{r^2}\partial_r\left(r^2{\bar\omega}e^{-2\nu}\right)U_l
\right] (l+1) Q_{l-1}Q_lY_{l-2}^m 
\Biggr\}
\ea
\label{om_r_th_form2}
\ee

\noindent
From $\Delta\omega_{\varphi r}=0$ we have,
\be
0 = \sum_l\Biggl\{
\ba[t]{l}
\biggl[
\partial_r\left[\frac{1}{r}\partial_r\left(r^2{\bar\omega}e^{-2\nu}\right)W_l\right]
-2l\partial_r\left({\bar\omega}e^{-2\nu}V_l\right)
\biggr] Q_{l+2}Q_{l+1}Y_{l+2}^m \\
 \\
+\biggl[
l\kappa\Omega\partial_r\left[e^{-2\nu}(h_l+U_l)\right]
-2m\partial_r\left({\bar\omega}e^{-2\nu}U_l\right)
+\frac{ml}{r^2}\partial_r\left(r^2{\bar\omega}e^{-2\nu}\right)U_l
\biggr] Q_{l+1} Y_{l+1}^m  \\
 \\
+\biggl[
	\ba[t]{l}
	m\kappa\Omega\partial_r\left(e^{-2\nu}V_l\right)
	+2\partial_r\left({\bar\omega}e^{-2\nu}V_l\right)
	\left((l+1)Q_l^2-l Q_{l+1}^2\right) \\
	\\
	+\frac{m^2}{r^2}\partial_r\left(r^2{\bar\omega}e^{-2\nu}\right)V_l
	+\partial_r\left[\frac{1}{r}\partial_r\left(r^2{\bar\omega}e^{-2\nu}
	\right)W_l\right] \left(Q_{l+1}^2+Q_l^2-1\right) \\
	\\
	+m\kappa\Omega e^{-2\nu}\left(\frac{16\pi r(\epsilon+p)}{l(l+1)}
					-\frac{1}{r}\right)e^{2\lambda}W_l
	\biggr] Y_l^m 
	\ea \\
 \\
-\biggl[
	\ba[t]{l}
	(l+1)\kappa\Omega\partial_r\left[e^{-2\nu}(h_l+U_l)\right] \\
	\\
	+2m\partial_r\left({\bar\omega}e^{-2\nu}U_l\right)
	+\frac{m(l+1)}{r^2}\partial_r\left(r^2{\bar\omega}e^{-2\nu}\right)U_l
	\biggr] Q_l Y_{l-1}^m  
	\ea \\
 \\
+\biggl[
\partial_r\left[\frac{1}{r}\partial_r\left(r^2{\bar\omega}e^{-2\nu}\right)W_l\right]
+2(l+1)\partial_r\left({\bar\omega}e^{-2\nu}V_l\right)
\biggr] Q_{l-1}Q_l Y_{l-2}^m 
\Biggr\}
\ea
\label{om_ph_r_form2}
\ee


Finally, 
let us rewrite these equations  using the orthogonality 
relation for spherical harmonics,
\be
\int Y_{l'}^{m'} Y_l^{m\ast} d\Omega = \delta_{ll'} \delta_{mm'},
\ee
where $d\Omega$ is the usual solid angle element on the unit 2-sphere.

\noindent
From $\Delta\omega_{\theta\varphi}=0$ we have, for all allowed $l$,
\be
0 = \ba[t]{l}
\left[ l(l+1)\kappa\Omega(h_l+U_l)-2m{\bar\omega}U_l\right] \\
 \\
+(l+1)Q_l \left[
\frac{e^{2\nu}}{r}\partial_r\left(r^2{\bar\omega}e^{-2\nu}\right)W_{l-1}
-2(l-1){\bar\omega}V_{l-1}
\right]  \\
 \\
-l Q_{l+1} \left[\frac{e^{2\nu}}{r}\partial_r
\left(r^2{\bar\omega}e^{-2\nu}\right)W_{l+1}
+2(l+2){\bar\omega}V_{l+1} \right]
\ea
\label{om_th_ph}
\ee

\noindent
Similarly, $\Delta\omega_{r\theta}=0$ leads to
\bea
0 &=&
(l-2)Q_{l-1}Q_l \left[
-2\partial_r\left({\bar\omega}e^{-2\nu}U_{l-2}\right)
+\frac{(l-1)}{r^2}\partial_r\left(r^2{\bar\omega}e^{-2\nu}\right)U_{l-2}
\right] 
\label{om_r_th} \\
&&\nn \\
&&+ Q_l \biggl[
	\ba[t]{l}
	(l-1)\kappa\Omega\partial_r\left(e^{-2\nu}V_{l-1}\right)
	-2m\partial_r\left({\bar\omega}e^{-2\nu}V_{l-1}\right) \\
	 \\
	+\frac{m(l-1)}{r^2}\partial_r\left(r^2{\bar\omega}e^{-2\nu}\right)V_{l-1}
	+(l-1)\kappa\Omega e^{-2\nu}\left(\frac{16\pi r(\epsilon+p)}{(l-1)l}
					-\frac{1}{r}\right)e^{2\lambda}W_{l-1}
	\biggr]
	\ea \nn \\
&&\nn \\
&&+\biggl[
	\ba[t]{l}
	m\kappa\Omega\partial_r\left[e^{-2\nu}(h_l+U_l)\right]
	+2\partial_r\left({\bar\omega}e^{-2\nu}U_l\right)
	\left((l+1)Q_l^2-l Q_{l+1}^2\right) \\
	 \\
	+\frac{1}{r^2}\partial_r\left(r^2{\bar\omega}e^{-2\nu}\right)U_l
	\left[m^2+l(l+1)\left(Q_{l+1}^2+Q_l^2-1\right)\right]
	\biggr]
	\ea \nn \\
&&\nn \\
&&- Q_{l+1} \biggl[
	\ba[t]{l}
	(l+2)\kappa\Omega\partial_r\left(e^{-2\nu}V_{l+1}\right)
	+2m\partial_r\left({\bar\omega}e^{-2\nu}V_{l+1}\right) \\
	 \\
	+\frac{m(l+2)}{r^2}\partial_r\left(r^2{\bar\omega}e^{-2\nu}\right)V_{l+1}
	+(l+2)\kappa\Omega e^{-2\nu}\left(\frac{16\pi r(\epsilon+p)}{(l+1)(l+2)}
					-\frac{1}{r}\right)e^{2\lambda}W_{l+1}
	\biggr]
	\ea \nn \\
&&\nn \\
&&+(l+3)Q_{l+1}Q_{l+2} \left[
2\partial_r\left({\bar\omega}e^{-2\nu}U_{l+2}\right)
+\frac{(l+2)}{r^2}\partial_r\left(r^2{\bar\omega}e^{-2\nu}\right)U_{l+2}
\right]
\nn
\eea

\noindent
and from $\Delta\omega_{\varphi r}=0$ we get
\bea
0 &=& Q_{l-1}Q_l \biggl[
\partial_r\left[\frac{1}{r}\partial_r
\left(r^2{\bar\omega}e^{-2\nu}\right)W_{l-2}\right]
-2(l-2)\partial_r\left({\bar\omega}e^{-2\nu}V_{l-2}\right)
\biggr] 
\label{om_ph_r} \\
&&\nn \\
&&+ Q_l \biggl[
	\ba[t]{l}
	(l-1)\kappa\Omega\partial_r\left[e^{-2\nu}(h_{l-1}+U_{l-1})\right] \\
	\\
	-2m\partial_r\left({\bar\omega}e^{-2\nu}U_{l-1}\right)
	+\frac{m(l-1)}{r^2}\partial_r\left(r^2{\bar\omega}e^{-2\nu}\right)U_{l-1}
	\biggr] 
	\ea \nn \\
&&\nn  \\
&&+\biggl[
	\ba[t]{l}
	m\kappa\Omega\partial_r\left(e^{-2\nu}V_l\right)
	+2\partial_r\left({\bar\omega}e^{-2\nu}V_l\right)
	\left((l+1)Q_l^2-l Q_{l+1}^2\right) \\
	\\
	+\frac{m^2}{r^2}\partial_r\left(r^2{\bar\omega}e^{-2\nu}\right)V_l
	+\partial_r\left[\frac{1}{r}\partial_r\left(r^2{\bar\omega}e^{-2\nu}
	\right)W_l\right] \left(Q_{l+1}^2+Q_l^2-1\right) \\
	\\
	+m\kappa\Omega e^{-2\nu}\left(\frac{16\pi r(\epsilon+p)}{l(l+1)}
					-\frac{1}{r}\right)e^{2\lambda}W_l
	\biggr]
	\ea \nn \\
&&\nn  \\
&&- Q_{l+1} \biggl[
	\ba[t]{l}
	(l+2)\kappa\Omega\partial_r\left[e^{-2\nu}(h_{l+1}+U_{l+1})\right] \\
	\\
	+2m\partial_r\left({\bar\omega}e^{-2\nu}U_{l+1}\right)
	+\frac{m(l+2)}{r^2}\partial_r\left(r^2{\bar\omega}e^{-2\nu}\right)U_{l+1}
	\biggr] 
	\ea \nn \\
&&\nn  \\
&&+ Q_{l+1}Q_{l+2} \biggl[
\partial_r\left[\frac{1}{r}\partial_r
\left(r^2{\bar\omega}e^{-2\nu}\right)W_{l+2}\right]
+2(l+3)\partial_r\left({\bar\omega}e^{-2\nu}V_{l+2}\right)
\biggr]
\nn
\eea

It is instructive to consider the Newtonian limit \cite{lf99}
\be
\om(r), \nu(r), \lambda(r), h_l(r) \rightarrow 0.
\label{Newt_lim}
\ee
of these perturbation equations.
We have already seen that Eq. (\ref{GR:sph_V}) is the relativistic 
generalization of the Newtonian mass conservation equation (LF,13)
(or Eq. (LF,42)), and that the other $O(1)$ perturbation equations, 
(\ref{GR:sph_H1}) and (\ref{GR:sph_h_0''}), simply vanish in the 
Newtonian limit. Similarly, one can readily observe that the 
conservation of vorticity equations have as their Newtonian limit 
the corresponding equations from Lockitch and Friedman \cite{lf99},
\[
\ba{lcl}
\mbox{Eq. (\ref{om_th_ph})} & \rightarrow  & \mbox{Eq. (LF,38)}, \\
\mbox{Eq. (\ref{om_r_th})}  & \rightarrow  & \mbox{Eq. (LF,40)}, \\
\mbox{Eq. (\ref{om_ph_r})} & \rightarrow  & \mbox{Eq.  (LF,39)}
\ea
\]
(and similarly for the other forms of these equations).

This correspondence leads us to expect the same structure for the 
relativistic modes as was found in the isentropic Newtonian case: we 
expect to find a discrete set of axial- and polar-led hybrid modes with 
opposite behavior under parity \cite{lf99}.  Further, we expect a 
one-to-one correspondence between these relativistic hybrid modes and the 
Newtonian modes, which the relativistic hybrids should approach in the 
Newtonian limit.

In deriving the components of the perturbed vorticity tensor in 
Newtonian gravity (LF,38)-(LF,40), no assumptions about the ordering 
of the perturbation variables $(\delta\rho,\delta v^a)$ in powers of 
the angular velocity $\Omega$ were required.  Thus, the theorem 
concerning the character of the Newtonian modes, [cf. Appendix A,
\cite{lf99}] applies to any discrete normal mode of a uniformly 
rotating isentropic star with arbitrary angular velocity.  

We conjecture that the perturbations of relativistic stars obey
the same principle: If $(\xi^\alpha, h_{\alpha\beta})$ with
$\xi^\alpha\neq 0$ is a discrete normal mode of a uniformly rotating 
stellar model obeying a one-parameter equation of state, then the 
decomposition of the mode into spherical harmonics $Y_l^m$ has 
$l=m$ as the lowest contributing value of $l$, when $m\neq 0$; and 
has 0 or 1 as the lowest contributing value of $l$, when $m=0$.

However, in deriving the curl of the perturbed Euler equation for 
relativistic models we have imposed assumptions that restrict its
generality.  We have derived Eqs. (\ref{om_th_ph})-(\ref{om_ph_r})
in a form that requires a slowly rotating equilibrium model, assumes 
the ordering (\ref{ordering}) and neglects terms of order $\Omega^2$
and higher.  Under these more restrictive assumptions, the following 
theorem holds.

\begin{thm}{Let $(g_{\alpha\beta}(\Omega), T_{\alpha\beta}(\Omega))$
be a family of stationary, axisymmetric spacetimes describing a 
sequence of stellar models in uniform rotation with angular velocity
$\Omega$ and obeying a one-parameter equation of state, where
$(g_{\alpha\beta}(0), T_{\alpha\beta}(0))$ is a static spherically
symmetric spacetime describing the non-rotating model.
Let $(\xi^\alpha(\Omega), h_{\alpha\beta}(\Omega))$ with 
$\xi^\alpha\neq 0$ be a family of discrete normal modes of these 
spacetimes obeying the same one-parameter equation of state, where
$(\xi^\alpha(0), h_{\alpha\beta}(0))$ is a stationary non-radial 
perturbation of the static spherical model. 
Let $(\xi^\alpha(\Omega_0), h_{\alpha\beta}(\Omega_0)$ be a member
of this family with $\Omega_0\ll \Omega_K$, the angular velocity of a 
particle in orbit at the star's equator. Then the decomposition of 
$(\xi^\alpha(\Omega_0), h_{\alpha\beta}(\Omega_0)$ into 
spherical harmonics $Y_l^m$ (i.e., into $(l, m)$ representations of 
the rotation group about its center of mass) has $l=m$ as the lowest 
contributing value of $l$, when $m\neq 0$; and $l=1$ as the lowest 
contributing value of $l$, when $m=0$.}
\label{thm2}
\end{thm}

We designate a 
non-axisymmetric\footnote{When $m=0$, there exists a set of modes
with parity $+1$ that may be designated as ``axial-led hybrids,''
since $\xi^\alpha$ and $h_{\alpha\beta}$ receive contributions only 
from axial terms with $l \ = \ 1, \ 3, \ 5,  \ \ldots$ and 
polar terms with $l \ = \ 2, \ 4, \ 6, \ \ldots$.}
mode with parity $(-1)^{m+1}$ an
``axial-led hybrid'' if $\xi^\alpha$ and $h_{\alpha\beta}$
receive contributions only from
\begin{center}
axial terms with $l \ = \ m, \ m+2, \ m+4,  \ \ldots$ and \\
polar terms with $l \ = \ m+1, \ m+3, \ m+5, \ \ldots$.
\end{center}
Similarly, we designate a 
non-axisymmetric\footnote{When $m=0$, there exists a set of modes
with parity $-1$ that may be designated as ``polar-led hybrids,''
since $\xi^\alpha$ and $h_{\alpha\beta}$ receive contributions only 
from polar terms with $l \ = \ 1, \ 3, \ 5, \ \ldots$ and
axial terms with $l \ = \ 2, \ 4, \ 6, \ \ldots$. The family of 
modes for which $\xi^\alpha$ and $h_{\alpha\beta}$ receive 
contributions only from polar terms with $l=0,2,4,\ldots$ and 
axial terms with $l=1,3,5,\ldots$ would have parity $+1$ and 
could be designated ``polar-led hybrids.''  However, these modes 
require a more general theorem to establish their character.}
mode with parity $(-1)^m$ a
``polar-led hybrid'' if $\xi^\alpha$ and $h_{\alpha\beta}$
receive contributions only from
\begin{center}
polar terms with $l \ = \ m, \ m+2, \ m+4,  \ \ldots$ and \\
axial terms with $l \ = \ m+1, \ m+3, \ m+5,  \ \ldots$.
\end{center}

We prove the theorem separately for each parity class in Appendix~C.

In essence, the theorem shows that if a mode of a slowly rotating 
isentropic star has a stationary non-radial perturbation as its 
spherical limit, then it is generically a hybrid mode with mixed 
axial and polar angular behavior.  An immediate consequence of the 
theorem is that the r-modes of isentropic stars (if they exist at all) 
must have $l=m$ (or $l=1$ if $m=0$), and it is well known that 
isentropic Newtonian stars retain a vestigial set of purely axial 
modes  - the ``classical r-modes'' - whose angular behaviour is a 
purely axial harmonic having $l=m$. Let us address the question of 
whether or not such pure r-mode solutions also exist in isentropic 
relativistic stars.

We apply the perturbation equations for isentropic stars to the case
of an axial mode belonging to a pure spherical harmonic of index $l$.  
In other words, let us assume that $h_l(r)$ and $U_l(r)$ (for some 
particular value of $l$) are the only non-vanishing coefficients in the 
spherical harmonic expansions (\ref{xi_exp}) and (\ref{h_components}) of 
the Lagrangian displacement, $\xi^\alpha$, and the perturbed metric, 
$h_{\alpha\beta}$, respectively.  

The set of equations to be satisfied are the zeroth order (spherical) 
equations (\ref{GR:sph_H1}), (\ref{GR:sph_h_0''}) and (\ref{GR:sph_V}) 
and the order $\Omega$ conservation of circulation equations 
(\ref{om_th_ph})-(\ref{om_ph_r}), as well as suitable boundary 
conditions at infinity and at the surface of the star, (Section \ref{sect4c}). 
Recall that as a result of Eq. (\ref{not_ind}), the conservation of 
circulation equations are linearly dependent and we need only satisfy 
two of them, say, Eqs. (\ref{om_th_ph}) and (\ref{om_r_th}).

With $h_l(r)$ and $U_l(r)$ the only non-vanishing perturbation variables,
Eqs. (\ref{GR:sph_H1}) and (\ref{GR:sph_V}) vanish identically, while Eq.
(\ref{GR:sph_h_0''}) remains unchanged.
Eq. (\ref{om_th_ph}) becomes,
\be
0 = \left[ l(l+1)\kappa\Omega(h_l+U_l)-2m{\bar\omega}U_l\right],
\label{koji2}
\ee
and Eq. (\ref{om_r_th}) with $l\rightarrow l+2$, $l\rightarrow l$ and
$l\rightarrow l-2$ gives the 
equations
\begin{eqnarray}
0 &=& 	lQ_{l+1}Q_{l+2} \left[
	-2\partial_r\left({\bar\omega}e^{-2\nu}U_l\right)
	+\frac{(l+1)}{r^2}\partial_r\left(r^2{\bar\omega}e^{-2\nu}\right)U_l
	\right], \label{koji3} \\
0 &=&	\ba[t]{l}
	m\kappa\Omega\partial_r\left[e^{-2\nu}(h_l+U_l)\right] \\
	\\
	+ 2\partial_r\left({\bar\omega}e^{-2\nu}U_l\right)
	\left((l+1)Q_l^2-l Q_{l+1}^2\right) \\
	 \\
	+\frac{1}{r^2}\partial_r\left(r^2{\bar\omega}e^{-2\nu}\right)U_l
	\left[m^2+l(l+1)\left(Q_{l+1}^2+Q_l^2-1\right)\right],
	\ea \label{koji4} \\
0 &=& (l+1)Q_{l-1}Q_l \left[
	2\partial_r\left({\bar\omega}e^{-2\nu}U_l\right)
	+\frac{l}{r^2}\partial_r\left(r^2{\bar\omega}e^{-2\nu}\right)U_l
	\right], \label{koji5}
\end{eqnarray}
respectively.  

Given the requirement that $l=m$ when $m>0$ (and $l=1$ when $m=0$), 
one readily finds that these equations can be satisfied only when $l=1$. 
{\it Thus, no purely axial modes with $l=m\geq 2$ exist 
in isentropic relativistic stars}~\cite{lo99}.   
The dipole ($l=1$) solutions turn out to be stationary ($\sigma=0$) 
and have the natural physical interpretation of a uniform rotation of 
the star\footnote{When $m=0$ and $l=1$, the solution corresponds to a 
small change in the angular velocity of the star about its original 
rotational axis.  When $l=m=1$, the solution represents uniform rotation 
of the star about an axis perpendicular to its original rotational axis. 
These solutions are derived in detail by Lockitch~\cite{lo99} and are a
generalization of the axial dipole modes studied in non-rotating 
relativistic stars by Campolattaro and Thorne \cite{ct70}.}.

In Newtonian isentropic stars there remained a large set of purely
axial modes with $l=m$; the $l=m=2$ mode being the one expected to
dominate the gravitational wave-driven instability of sufficiently 
hot and rapidly rotating neutron stars \cite{lom98,aks98}.  
In isentropic relativistic stars, however, we see that all such
pure r-modes with $l=m\geq 2$ are forbidden by the perturbation
equations, and instead must be replaced by axial-led hybrids. 
We explicitly construct these important hybrid modes to first
post-Newtonian order in Section \ref{sect5c}.

\subsection{Boundary Conditions}
\label{sect4c}

Having understood the general nature of the relativistic perturbation 
problem and derived the relevant perturbation equations for both
isentropic and non-isentropic stars, 
we want to determine mode-solutions. 
Before we can do this, we need to discuss the boundary conditions that
should be imposed.

For non-isentropic stars, the zeroth order variables are governed by 
the single equation (\ref{singeq}), while for isentropic stars we have 
the set of perturbation equations (\ref{GR:sph_H1}), (\ref{GR:sph_h_0''}),
(\ref{GR:sph_V}) and (\ref{om_th_ph})-(\ref{om_ph_r}).
A physically reasonable solution $(\xi^\alpha, h_{\alpha\beta})$
to these equations must be regular everywhere in the spacetime.  
Of course, the fluid 
variables $W_l(r)$, $V_l(r)$ and $U_l(r)$ (for all $l$) have support 
only inside the star, $r\in [0,R]$.  The metric functions $H_{1,l}(r)$ 
will also have support only inside the star (for all $l$), since they 
are directly proportional to $W_l(r)$ by Eq. (\ref{GR:sph_H1}).
The metric functions $h_l(r)$, on the other hand, satisfy a nontrivial 
differential equation, (\ref{GR:sph_h_0''}), in the exterior spacetime
and will, therefore, have support on the whole domain $r\in [0,\infty]$.
Let us now consider the boundary and matching conditions that our
solutions must satisfy.

At the surface of the star, $r=R$, the perturbed pressure, $\Delta p$, 
must vanish. (This is how one defines the surface of the perturbed star.)
The Lagrangian change in the pressure is given by 
Eq. (\ref{Del_etc}),
\be
\Delta p = -\half \, \Gamma_1 \, p \, 
q^{\alpha\beta}\Delta g_{\alpha\beta}.
\ee
Making use of Eq. (\ref{form_Dp}) and the equilibrium equations
(\ref{GR_sph:tov}) and (\ref{GR_sph:dnudr}), we find that at $r=R$
\be
0 = \Delta p = \frac{-\ep M_0}{R^2\left(R-2M_0\right)}
\sum_l W_l(R) Y_l^m e^{i\sigma t}
\label{bc}
\ee
where $M_0=M(R)$ is the gravitational mass of the equilibrium star and
satisfies $2M_0<R$.

For the equations of state we 
consider\footnote{This restriction can be dropped if the boundary 
condition $\Delta p(r=R)=0$ is replaced by $\Delta h(r=R)=0$, with the 
comoving enthalpy $h\equiv \int_0^p dp'/(\ep(p')+p')$.}
the energy density $\epsilon(r)$ either goes to a constant or vanishes 
at the surface of the star in the manner (this would be the 
behavior for a polytrope),
\be
\ep(r) \sim \left(1-\frac{r}{R}\right)^k
\label{EOStype}
\ee
(for some constant $k$). 
In both cases, it is required that 
\be
W_l(R)=0 \ \ \ \ \mbox{(all $l$)}.
\label{bc_on_W}
\ee
If $\epsilon(R)\neq 0$, then Eq. (\ref{bc}) requires this directly.
On the other hand, if $\ep$ vanishes as in Eq. (\ref{EOStype}) then
$V_l(r)$ will diverge at the surface by Eq. (\ref{GR:sph_V}) if
(\ref{bc_on_W}) is not satisfied.  By Eq. 
(\ref{GR:sph_H1}), this also implies that $H_{1,l}(r)$ vanishes 
at the surface of the star. This boundary condition is 
(obviously) relevant only
in the isentropic case.

In the exterior vacuum spacetime, $r>R$, we have only to satisfy the 
single 
equation (\ref{GR:sph_h_0''}) for all $l$, which becomes
\be
h_l^{''} + \left[ \frac{(2-l^2-l)}{r^2}e^{2\lambda} 
	- \frac{2}{r^2} \right] h_l  = 0,
\ee
or
\be
(1-\frac{2M_0}{r}) h_l^{''} - \left[ \frac{l(l+1)}{r^2} 
	- \frac{4M_0}{r^3} \right] h_l  = 0,
\label{h_l''_ext}
\ee
where we have used $e^{-2\lambda}=(1-2M_0/r)$ for $r>R$.

Since this exterior equation does not couple $h_l(r)$ for different 
values of $l$, we can find its solution explicitly.  The solution that 
is regular at spatial infinity can be written
\be
h_l(r) = \sum_{s=0}^\infty {\hat h}_{l,s} 
\left(\frac{R}{r}\right)^{l+s}.
\label{h_ext}
\ee
If we substitute this series expansion into Eq. (\ref{h_l''_ext}), 
we find the following recursion relation for the expansion 
coefficients,
\be
{\hat h}_{l,s} = \left(\frac{2M_0}{R}\right) 
\frac{(l+s-2)(l+s+1)}{s(2l+s+1)} {\hat h}_{l,s-1}
\label{h_ext_soln}
\ee
with ${\hat h}_{l,0}$ an arbitrary normalization constant.  We, 
therefore, have the full solution to zeroth order in $\Omega$ of the 
perturbation equations in the exterior spacetime.

This exterior solution must be matched at the surface of the star to 
the interior solution for $h_l(r)$.  One requires that the solutions
be continuous at the surface,
\be
\lim_{\varepsilon\rightarrow 0}  \left[
h_l(R-\varepsilon) - h_l(R+\varepsilon) \right] = 0,
\label{cont_cond}
\ee
for all $l$, and that the Wronskian of the interior and exterior 
solutions vanish at $r=R$, i.e. that
\be
\lim_{\varepsilon\rightarrow 0} \left[
h_l(R-\varepsilon) h'_l(R+\varepsilon) 
- h'_l(R-\varepsilon) h_l(R+\varepsilon)
\right] = 0,
\label{match_cond}
\ee
for all $l$.

Thus, in solving the perturbation equations to zeroth order in $\Omega$ 
we need only work in the interior of the star (as in the Newtonian case).  
In the interior of a non-isentropic star, the perturbation 
$(\xi^\alpha, h_{\alpha\beta})$ must only satisfy Eq. (\ref{singeq}) 
together with the matching conditions (\ref{cont_cond}) and 
(\ref{match_cond}).  In the isentropic case we have the full 
set of coupled equations (\ref{GR:sph_H1}), (\ref{GR:sph_h_0''}), 
(\ref{GR:sph_V}) and (\ref{om_th_ph})-(\ref{om_ph_r}) for all $l$, subject 
to the boundary and matching conditions (\ref{bc_on_W}), (\ref{cont_cond}) 
and (\ref{match_cond}).

Finally, we note that since we are working in linearized perturbation
theory there is a scale invariance to the equations.  If 
$(\xi^\alpha, h_{\alpha\beta})$ is a solution to the perturbation
equations then $(K\xi^\alpha, Kh_{\alpha\beta})$ is also a 
solution, for constant $K$. This will not affect the non-isentropic
calculation, but in the isentropic case we will find it convenient to 
impose the following normalization condition in addition to the 
boundary and matching conditions just discussed;
\be
\begin{array}{ll}
U_m(r=R) = 1 & \mbox{for axial-hybrids,} \\
U_{m+1}(r=R) = 1 & \mbox{for polar-hybrids.} 
\end{array}
\label{norm_cond}
\ee

\section{Relativistic corrections to the r-modes
of uniform density stars}
\label{sect5}

In a future paper, we will consider the general problem of numerically 
solving for the r-modes and hybrid modes of fully relativistic stars. 
Preliminary results have already been presented by Lockitch \cite{lo99}.
For now, we will focus on the post-Newtonian corrections to the 
Newtonian r-modes.  The equilibrium structure of a slowly rotating star 
with uniform density is particularly simple~\cite{cm74}
and lends itself readily to such a post-Newtonian analysis.
The results we obtain in this way provide important insights into the
relativistic corrections to the familiar Newtonian r-modes.


\subsection{A post-Newtonian uniform density model}
\label{sect5a}

For a spherically symmetric star with constant density,
\be
\ep(r)=\frac{3M_0}{4\pi R^3},
\label{GR:equil_ep}
\ee
the equilibrium equations (\ref{GR_sph:eos})-(\ref{GR_sph:mass_def}) have 
the well-known exact solution inside the star ($r\leq R$),
\bea
p(r) &=& \ep\left\{
\frac{\left(1-\tmr\right)^\half 
- \left[1-\tmr\left(\rx\right)^2\right]^\half}
{3\left[1-\tmr\left(\rx\right)^2\right]^\half 
- \left(1-\tmr\right)^\half}
\right\}
\label{GR:equil_p} \\
M(r) &=& M_0 \left(\frac{r}{R}\right)^3 \\
e^{2\nu(r)} &=& \left\{
\frac{3}{2}\left[1-\tmr\left(\rx\right)^2\right]^\half 
- \frac{1}{2}\left(1-\tmr\right)^\half
\right\}^{2} \\
e^{-2\lambda(r)} &=& 1 - \frac{2M_0}{R}\left(\frac{r}{R}\right)^2
\eea
where $M_0$ is the gravitational mass of the star and $R$ is its radius.

To find the equilibrium solution corresponding to a slowly rotating
star, we must also solve Hartle's~\cite{h67} equation (\ref{hartle}),
\be
0 = r^2\bom'' + [4-r(\nu'+\lambda')]r\bom'-4r(\nu'+\lambda')\bom
\label{hartle2}
\ee 
(see also \cite{cm74}) where we may use the spherical solution to write 
\bea
r(\nu'+\lambda')
 &=& 4\pi r^2(\ep+p)e^{2\lambda} \\
 &=& \frac{3\left(\tmr\right)\left(\rx\right)^2 
\left(1-\tmr\right)^\half}
{\left[1-\tmr\left(\rx\right)^2\right]\left\{
3\left[1-\tmr\left(\rx\right)^2\right]^\half 
- \left(1-\tmr\right)^\half
\right\}}. \nn
\eea

To simplify the problem, we expand our equilibrium solution in powers
of $(2M_0/R)$ and work only to linear 
order\footnote{This expansion will give us the first post-Newtonian
(1PN) corrections to the Newtonian r-modes.}. 
We will need the expressions,
\be
r(\nu'+\lambda') = \frac{3}{2}\left(\rx\right)^2\left(\tmr\right)
+ O\left(\tmr\right)^2
\label{GR:equil_nulam}
\ee
and
\be
e^{-2\nu} = 1+\left[\frac{3}{2}
-\frac{1}{2}\left(\rx\right)^2\right]\left(\tmr\right)
+ O\left(\tmr\right)^2
\label{GR:equil_gtt}
\ee

Since we are also working to linear order in the star's angular 
velocity, we may set $\Omega=1$ without loss of generality. 
We write,
\be
\bar\omega =
\sum^\infty_{i=0} \omega_i \left(\frac{r}{R}\right)^{2i}
\ee
and solve Eq. (\ref{hartle2}) subject to the boundary condition 
(\ref{om_cond}) at the surface of the star,
\be
1 = \Omega = \left[\bom + \frac{1}{3}R\ \bom'\right]_{r=R}
\label{hartleBC}
\ee
To order $(2M_0/R)$ the solution is,
\be
\bom(r) = 1 - \left(1-\frac{3r^2}{5R^2}\right)\left(\tmr\right)
+ O\left(\tmr\right)^2.
\label{GR:equil_bom}
\ee

\subsection{Non-isentropic stars}
\label{sect5b}

In order to find the find the relativistic analogue
to the familiar Newtonian r-modes of non-isentropic stars, 
we insert the above expressions in (\ref{singeq}).
We also assume that the mode-frequency can be approximated as
\be
\kappa = \frac{2m}{l(l+1)}\left[1+\kappa_1\left(\tmr\right)
+ O\left(\tmr\right)^2 \right] \ .
\ee
and that the eigenfunction takes the form
\begin{equation}
h_{l} \approx h_l^{(0)}(r) \left(\tmr\right)
+ O\left(\tmr\right)^2 \ .
\end{equation}
Solutions of this form would then lead to $U_l \sim O(1)$
via (\ref{hUrel}). 

Under these assumptions, (\ref{singeq}) is  trivially satisfied 
to leading order. At order $(2M_0/R)^2$ we find an equation 
\begin{equation}
\left[ \kappa_1 + 1 - {3r^2 \over 5R^2} \right]
\left\{ h_l^{(0)\prime \prime} - {l(l+1) \over r^2} h_l^{(0)} \right\}
+ {6\over  R^2} h_l^{(0)}
=0
\label{first}\end{equation}
Before proceeding, it is useful to compare our definition of 
the post-Newtonian eigenvalue $\kappa_1$ to 
the  eigenvalue $\alpha$ we used in Section \ref{sect4a}. We then 
immediately see that
\begin{equation}
\alpha = 1 + \kappa_1 \left( \tmr \right)  
\end{equation}
and deduce that the established range
for possible eigenfrequencies translates into
\begin{equation}
-1 < \kappa_1 <  0 \ .
\end{equation}
Within this range there are two possibilities. If $\kappa_1 \le -2/5$ 
we will have a singular eigenvalue problem, while for $-2/5< \kappa_1 < 0$
the problem is non-singular.
To determine the r-modes to first order in $2M_0/R$ for the uniform
density model, one need only consider the simpler non-singular
situation, because the eigenvalues of the relativistic r-modes turn out
to be in the nonsingular range.  The continuous part of the spectrum
\cite{k98,bk99}, as noted earlier, may be an artifact of an
approximation in which the frequency is real.

\begin{figure}
\begin{center}
\includegraphics{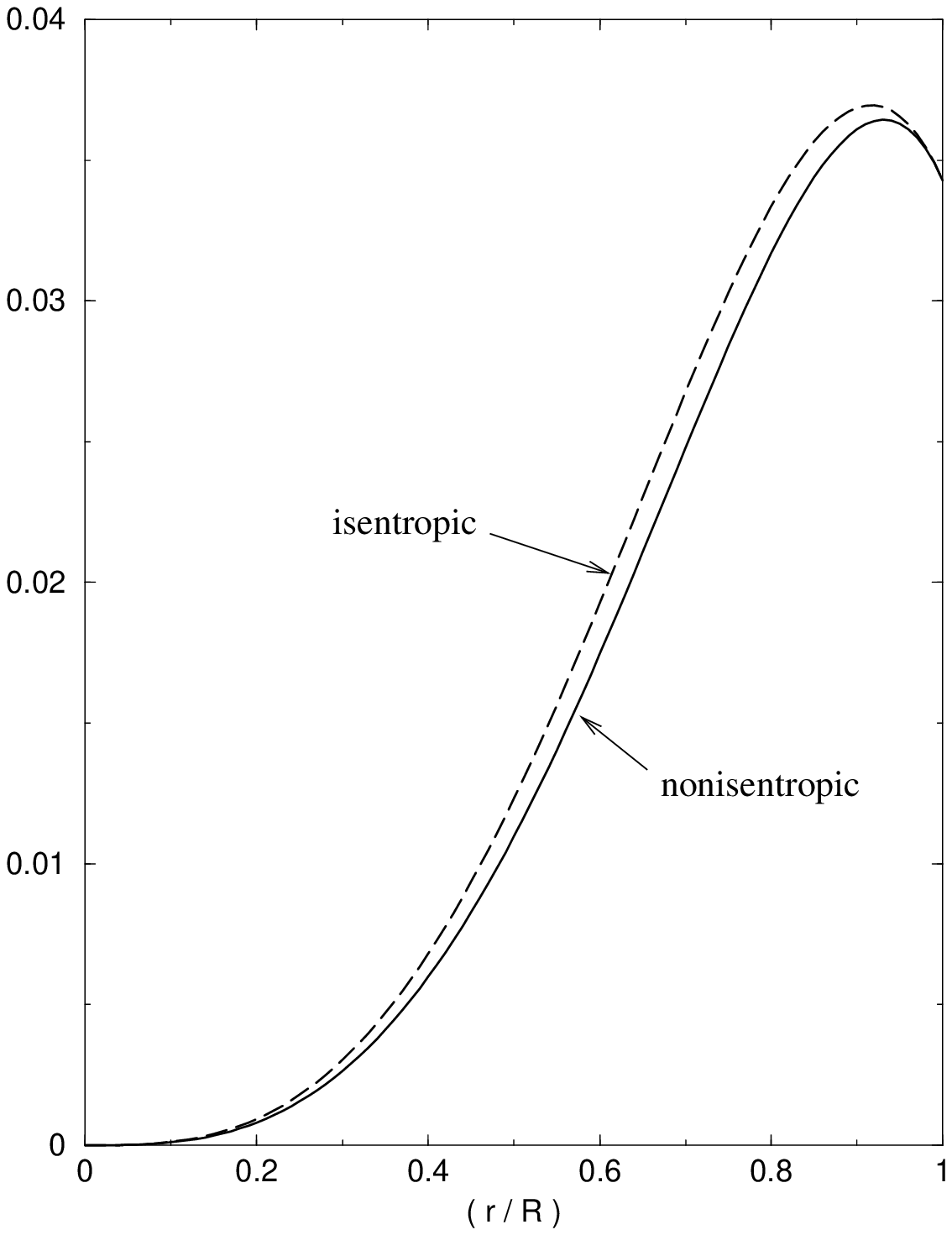}
\end{center}
\caption{Numerically determined post-Newtonian r-mode eigenfunction 
$h_l(r)$ for $l=2$ in the interior of a uniform density 
star (solid line). The result is compared to the corresponding 
eigenfunction for the particular hybrid mode that is the 
relativistic counterpart of the Newtonian $l=m=2$ r-mode in an 
isentropic star (shown as a dashed curve).  Of course, in the 
isentropic case several other functions (such as $W_{m+1}$ and 
$V_{m+1}$) are also non-zero (see Fig. 2).  The functions are
normalized to agree at the surface of the star and the overall
scale is set by the normalization of Fig.~2.}
\label{fig1}
\end{figure}

We can rewrite (\ref{first}) as
\begin{equation}
{d^2 h_l^{(0)} \over dr^2} - \left[ {l(l+1) \over r^2} +
{30 \over 3r^2-5R^2(\kappa_1+1)} \right] h_l^{(0)} = 0
\label{nonsing}\end{equation}
As long as $-2/5< \kappa_1 < 0$ this 
equation can readily be integrated, and the solutions are
well-behaved at all values of $r$. 
We have integrated (\ref{nonsing}) using a fourth-order
Runge-Kutta scheme, initiated from the appropriate regular
power series solution close to the centre of the star. That is, we 
use 
\begin{equation}
h_{l}^{(0)} \approx D r^{l+1} \left\{ 1 - { 6 r^2 \over 
R^2 (\kappa_1+1)[(l+2)(l+3)-l(l+1)]} \right\}
\end{equation}  
at an initial point close to $r=0$ and then integrate (\ref{nonsing})
to the surface $r=R$. At the surface we demand that $h_{l}^{(0)}$ and its 
derivative can be smoothly matched to the exterior solution
according to (\ref{match_cond}). For each value of $l$ we then find a single
acceptable solution, corresponding to a distinct eigenvalue $\kappa_1$.
These eigenvalues, for $l=2-10$, are listed in Table~\ref{tab1}.
It should be recalled that the tabulated eigenvalues
correspond to mode-frequencies (in the inertial frame) given 
by
\begin{equation}
\sigma \approx - m\Omega + {2m\Omega \over l(l+1)} \left[ 1 + \kappa_1\left(\tmr\right)
\right]  + O \left( \tmr \right)^2
\end{equation}
A typical eigenfunction, corresponding to $l=2$, is shown in 
Fig.~\ref{fig1}.

We thus find
a single post-Newtonian 
r-mode solution for each allowed combination of $l$ and $m$. 
This is very much in accordance with the Newtonian r-mode
results for non-isentropic stars at order $\Omega$ (the degeneracy
of these modes is not broken until at order $\Omega^2$).
The main difference in the relativistic case 
is that the post-Newtonian corrections (of order $2M_0/R$) break
the degeneracy at order $\Omega$ and make
it possible for us to determine the eigenfunctions.
 
Given these results we expect similar r-mode
solutions to exist also in the fully relativistic case. 
It is, in fact, easy to demonstrate this and
we have extended our calculation for uniform density star to include
all terms in (\ref{singeq}). We then find that the mode-eigenvalue is 
always such that $\alpha-\tilde{\omega}=\alpha-\bar{\omega}/\Omega\neq 0$ 
in the interior
of the star (recall the discussion in Section \ref{sect4a}). 
The solutions to the problem are thus regular. The
associated eigenvalues, for $l=2-10$ and a star with compactness
$M/R=0.2$ are given, 
and compared to the post-Newtonian results, in Table~\ref{tab1}.  

\subsection{Isentropic stars}
\label{sect5c}


Having established that discrete r-mode solutions exist for 
non-isentropic relativistic stars we now turn to the isentropic case.
As we have shown, we will then not have purely axial 
solutions (for $l\ge 2$). Instead, we need to 
calculate hybrid modes by  solving Eqs. (\ref{GR:sph_H1}), 
(\ref{GR:sph_h_0''}), (\ref{GR:sph_V}) and (\ref{om_th_ph})-(\ref{om_ph_r})
subject to the boundary, matching and normalization conditions
(\ref{bc_on_W}), (\ref{cont_cond}), (\ref{match_cond}) and (\ref{norm_cond}). 
As in the non-isentropic case, 
we seek the post-Newtonian corrections to the well-known Newtonian 
r-modes. For isentropic stars such modes exist only for $l=m$ with
frequency and radial dependence given by,
\bea
\kappa &=& \frac{2}{(m+1)} \\ \nn \\
U_m &=& \left(\frac{r}{R}\right)^{m+1}.
\eea
Therefore, let us make the following Ansatz for our perturbed solution
inside the star. We assume that the coefficients of the spherical harmonic 
expansions (\ref{xi_exp}) and (\ref{h_components}) of the Lagrangian 
displacement, $\xi^\alpha$, and the perturbed metric, $h_{\alpha\beta}$, 
respectively, have the form,
\bea
\kappa &=& \frac{2}{(m+1)}\left[1+\kappa_1\left(\tmr\right)
+ O\left(\tmr\right)^2 \right]
\label{ans:freq} \\ \nn \\
U_m(r) &=& \left(\rx\right)^{m+1}\left[
1+u_{m,0}\left(1-\frac{r^2}{R^2}\right)\left(\tmr\right)
+ O\left(\tmr\right)^2\right]
\label{ans:u_m} \\ \nn \\
h_m(r) &=& \left(\rx\right)^{m+1}
\left[h_{m,0} + h_{m,1}\left(\rx\right)^2\right]\left(\tmr\right)
+ O\left(\tmr\right)^2
\label{ans:v_m+1} \\ \nn \\
W_{m+1}(r) &=& w_{m,0}\left(\rx\right)^{m+1}
\left(1-\frac{r^2}{R^2}\right)\left(\tmr\right) + O\left(\tmr\right)^2
\label{ans:h_m} \\ \nn \\
V_{m+1}(r) &=& \left(\rx\right)^{m+1}
\left[v_{m,0} + v_{m,1}\left(\rx\right)^2\right]\left(\tmr\right)
+ O\left(\tmr\right)^2
\label{ans:w_m+1} \\ \nn \\
U_{m+2}(r) &=& u_{m+2,0}\left(\rx\right)^{m+3}\left(\tmr\right)
+ O\left(\tmr\right)^2
\label{ans:u_m+2}
\eea
where $\kappa_1$, $u_{m,0}$, $h_{m,0}$, $h_{m,1}$, $w_{m+1,0}$, $v_{m+1,0}$, 
$v_{m+1,1}$ and $u_{m+2,0}$ are (as yet) unknown constants.  
We have chosen the form of $U_m(r)$ so as to automatically
satisfy the normalization condition (\ref{norm_cond}) and we have 
chosen the form of $W_{m+1}(r)$ so as to automatically satisfy the boundary
condition (\ref{bc_on_W}).  Note that we have assumed that 
$h_l$, $V_{l'}$, $W_{l'}$ and $U_{l''}$ are of order $(2M_0/R)^2$ 
or higher for all $l>m$, $l'>m+1$ and $l''>m+2$.  We will justify 
this Ansatz by showing self-consistently that such a solution satisfies 
the perturbation equations.

Observe that the exterior solution (\ref{h_ext}) for $h_m(r)$ 
already has a natural expansion in powers of $(2M_0/R)$ as a 
result of the recursion relation (\ref{h_ext_soln}),
\be
h_m(r) = {\hat h}_{m,0} \left(\frac{R}{r}\right)^m\left(\tmr\right)
+ O\left(\tmr\right)^2.
\ee
The normalization constant, ${\hat h}_{m,0}$, is determined by the 
matching condition (\ref{cont_cond}),
\be
{\hat h}_{m,0} = h_{m,0} + h_{m,1} 
\ee
while (\ref{match_cond}) imposes the following condition on the interior
solution,
\be
0 = {\hat h}_{m,0} \biggl\{-m(h_{m,0}+h_{m,1}) 
- \left[ (m+1)h_{m,0} + (m+3)h_{m,1} \right]
\biggr\}
\ee
or,
\be
0 = (2m+1)h_{m,0} + (2m+3)h_{m,1}
\label{GR:ex_h_BC}
\ee

We turn now to the isentropic perturbation equations, (\ref{GR:sph_H1}),
(\ref{GR:sph_h_0''}), (\ref{GR:sph_V}) and 
(\ref{om_th_ph})-(\ref{om_ph_r}). Recall that these latter three
equations are not linearly independent, being related by Eq. 
(\ref{not_ind}).  Also, because Eq. (\ref{GR:sph_H1}) merely 
expresses $H_{1,l}(r)$ in terms of $W_l(r)$, we may eliminate
$H_{1,l}(r)$ from our system and ignore Eq. (\ref{GR:sph_H1}). 
Thus, a complete set of perturbation equations is provided by 
Eqs. (\ref{GR:sph_h_0''}), (\ref{GR:sph_V}), 
(\ref{om_th_ph}) and (\ref{om_r_th}) for all allowed values of $l$.

We expand these equations to first post-Newtonian order using
Eqs.  (\ref{GR:equil_nulam}), (\ref{GR:equil_gtt}) and 
(\ref{GR:equil_bom}) to replace the equilibrium quantities and 
using our Ansatz, Eqs. (\ref{ans:freq})-(\ref{ans:u_m+2}) to 
replace the various perturbation variables.  The result is an
algebraic system of seven independent equations, which, together with 
our matching condition (\ref{GR:ex_h_BC}) allows us to uniquely find
our eight unknown constants, $\kappa_1$, $u_{m,0}$, $h_{m,0}$, 
$h_{m,1}$, $w_{m+1,0}$, $v_{m+1,0}$, $v_{m+1,1}$ and $u_{m+2,0}$.
These equations are derived in detail by Lockitch\cite{lo99}. Here, we
will simply present the resulting solution,


\be
\kappa = \frac{2}{(m+1)}\left[1
-\frac{4(m-1)(2m+11)}{5(2m+1)(2m+5)}\left(\tmr\right)
+ O\left(\tmr\right)^2 \right]
\label{GR:ex_sol:freq}
\ee

\be
U_m(r) = \left(\rx\right)^{m+1}\left[
1+u_{m,0}\left(1-\frac{r^2}{R^2}\right)\left(\tmr\right)
+ O\left(\tmr\right)^2\right]
\label{GR:ex_sol:u_m}
\ee

\be
h_m(r) = \left(\rx\right)^{m+1}\left[-\frac{3}{(2m+1)} 
+ \frac{3}{(2m+3)}\left(\rx\right)^2\right]\left(\tmr\right)
+ O\left(\tmr\right)^2
\label{GR:ex_sol:v_m+1}
\ee

\be
W_{m+1}(r) = (m+1)(m+2)K\left(\rx\right)^{m+1}
\left(1-\frac{r^2}{R^2}\right)\left(\tmr\right) + O\left(\tmr\right)^2
\label{GR:ex_sol:h_m}
\ee

\be
V_{m+1}(r) = K\left(\rx\right)^{m+1}
\left[(m+2) - (m+4)\left(\rx\right)^2\right]\left(\tmr\right)
+ O\left(\tmr\right)^2
\label{GR:ex_sol:w_m+1}
\ee

\be
U_{m+2}(r) = -K\, Q_{m+2}\frac{(m+1)^2(m+3)}{(2m+3)}
\left(\rx\right)^{m+3}\left(\tmr\right)
+ O\left(\tmr\right)^2
\label{GR:ex_sol:u_m+2}
\ee
where we have defined the constant
\be
K \equiv \frac{6(m-1)Q_{m+1}}{5(m+2)(2m+5)}
\ee
and where
\be
u_{m,0} = - \frac{K\, Q_{m+1}}{24m(m+2)(2m+3)}
\ba[t]{l}
\biggl\{48(m+1)^4(m+3)^2 \\
 \\
+(2m+3)^2(2m+5)\biggl[m(m+2)^2-48\biggr]
\biggr\}
\ea
\ee

Since our solution satisfies the full perturbation equations 
to order $(2M_0/R)$, our Ansatz was self-consistent.  Thus, 
we have explicitly found the first post-Newtonian corrections 
to the $l=m$ Newtonian r-modes of isentropic uniform density stars.

\begin{figure}
\begin{center}
\resizebox{5.0in}{!}{\includegraphics{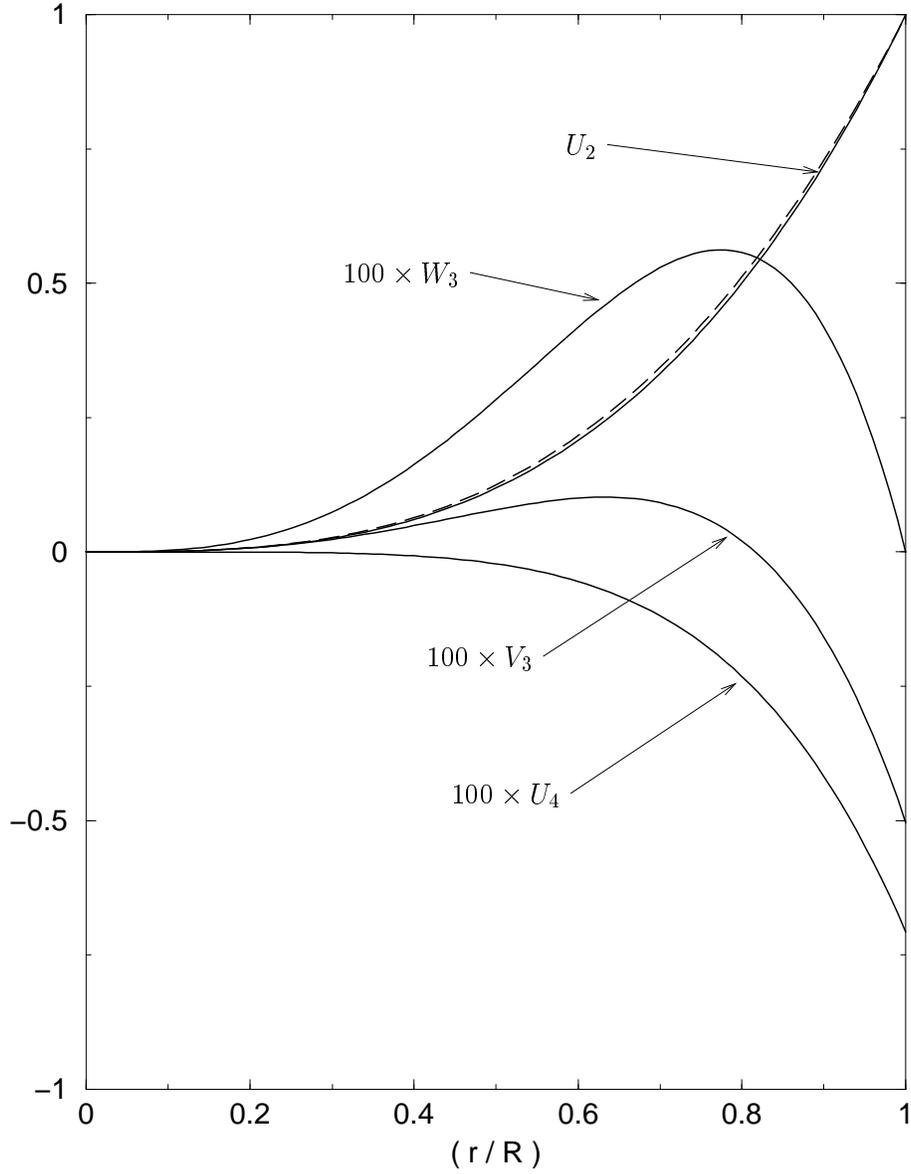}}
\end{center}
\caption{The $(r/R)^3$ radial dependence of the Newtonian $l=m=2$ r-mode 
is shown (dashed curve) together with the post-Newtonian
corrections to this mode for a uniform density star of 
compactness $2M/R=0.2$, i.e., the coefficients $U_l(r)$, 
$W_l(r)$, and $V_l(r)$ with $l\leq 4$ of the spherical harmonic 
expansion (\ref{xi_exp}) (solid curves).  The vertical scale 
is set by the normalization of $U_2(r)$ to unity at the surface 
of the star, and the other coefficients have been scaled by a 
factor of 100. Thus, while the relativistic corrections to the 
equilibrium structure of the star are of order $20\%$, the 
relativistic corrections to the r-mode are only of order $1\%$.}
\label{fig2}
\end{figure}

The solution reveals the expected mixing of axial and polar 
terms in the spherical harmonic expansion of $\xi^\alpha$.
All of the isentropic Newtonian r-modes with $m\geq 2$ pick up 
both axial and polar 
corrections\footnote{When $m=1$, the constant $K$ vanishes and 
we recover the axial dipole solution mentioned in Section \ref{sect4b}.} 
of order $(2M_0/R)$, becoming axial-led hybrid modes
of the relativistic star. The $l=m=2$ hybrid mode is shown in 
Figs.~\ref{fig1} and \ref{fig2}, and compared to the corresponding
r-mode in a non-isentropic star (see Section \ref{sect5b}).  

In addition, we see from Eq. (\ref{GR:ex_sol:freq}) that the 
Newtonian r-mode frequency also picks up a small relativistic 
correction. The frequency decreases, just as it does in the 
non-isentropic case (see Table~\ref{tab1}), and it is natural that 
general relativity will have such an effect. One reason is that
gravitational redshift will tend to decrease the fluid oscillation 
frequencies measured by a distant inertial observer. Also, because
these modes are rotationally restored they will be affected by the
dragging of inertial frames induced by the star's rotation.
The Coriolis force is ``determined not by the angular velocity 
$\Omega$ of the fluid relative to a distant observer but by its 
angular velocity relative to the local inertial frame, 
$\bar\om(r)$.'' (Hartle and Thorne~\cite{ht68})  
Thus, the Coriolis force decreases - 
and the modes oscillate less rapidly - as the dragging of inertial 
frames becomes more pronounced.  

Finally, we note that the metric perturbation (whose radial 
behaviour is determined by the function $h_m$) is of the same 
order as the post-Newtonian corrections to the fluid perturbation. 
Thus, there is no justification for the Cowling approximation in
constructing the hybrid mode solutions.  In Newtonian theory, 
the Cowling approximation 
corresponds to neglecting the variation in the gravitational
potential. The original motivation for this \cite{c41} is that 
some pulsation modes (in particular the g-modes) are mainly located
in the less dense regions close to the surface of the star, and
do not involve large mass motion. Hence, they will lead to 
variations in the gravitational potential that are small compared to 
the associated fluid velocities. The obvious generalization 
of this approach to general relativity would be to
discard all metric perturbations \cite{mcd83}. However, as was pointed
out by Finn \cite{fi88}, this approximation is not  natural
for relativistic g-modes. The main reason is that, even though these
modes involve small density perturbations they could involve
large fluid velocities. Hence, 
Finn argues that one should keep those 
metric perturbations that can be associated with ``momentum
transport'' in calculations of g-modes.  As is easy to see,
similar arguments can be used for the modes
we consider in the present paper. This would suggest
that one should not discard the metric perturbations 
$h_1$, $h_0$ and $H_1$ in 
the relativistic Cowling approximation for 
r-modes and hybrid modes. Interestingly, should we adopt this
point of view we retain the main perturbation equations we have 
used in the present paper. Hence, this ``approximation'' would
be  consistent with our results. Furthermore, 
this would explain why the attempts to find relativistic
r-modes within the Cowling approximation (by neglecting all 
metric perturbations) have failed \cite{kh99}.
Of course, this discussion has little relevance for 
the present study. But it could be of crucial importance
for attempts to find r-modes in numerical simulations (by studying
fluid motion in relativistic simulations with a ``frozen'' metric)
that are currently under way \cite{font,sf00}.


\section{Discussion}

In this paper we have taken the first steps towards an understanding
of both r-modes and rotational hybrid modes of rotating relativistic stars. 
We have derived the perturbation equations that govern these modes
to linear order in the rotation frequency $\Omega$
(at which the star is still spherical). For non-isentropic
stars we have focused on modes that have a purely axial 
limit as $\Omega\to 0$. These would be a natural
relativistic generalization of the Newtonian r-modes. 
For isentropic stars (and multipoles $l\ge 2$) we have shown
that no such modes exist in the relativistic case, 
even though Newtonian stars retain a vestigial set 
corresponding to $l=m$. Instead, all modes of isentropic
stars must have a hybrid nature. Having derived the relevant 
perturbation equations we 
calculate relativistic corrections at the first post-Newtonian
level (order $2M_0/R$) to the Newtonian r-modes
of both non-isentropic and isentropic stars.

It is worth pointing out that, even though our results 
for isentropic and non-isentropic stars are quite different,
the particular modes that we have focussed on (the analogues of
the vestigial $l=m$ r-modes that remain for isentropic
Newtonian stars)
are not too dissimilar. As is clear from the results given in 
Table~\ref{tab1}, the mode frequencies in the two cases we have considered
do not differ by more than a few percent. Furthermore, we can see from 
Figure~\ref{fig1} that the axial eigenfunctions $h_{0,l}$ are similar. 
There are, of course, still considerable differences between the two cases. 
In the non-isentropic case we predict that purely axial modes exist for all
combinations of $l$ and $m\neq0$, while in the isentropic case
all modes are hybrids. Still, the fact that our results for the 
two cases seem consistent is encouraging. We anticipate that further
work will eventually unveil a behavior quite similar to that of 
the Newtonian problem, for which the detailed isentropic limit 
has been investigated by Yoshida and Lee~\cite{yosh00}. 

This paper represents progress in several 
important directions, but
a considerable amount of work remains before we
can claim to have a complete understanding of the nature
of the r-modes and hybrid modes in relativity. 
For example, we have not yet 
discussed how the inferred
changes in both mode-frequency and eigenfunction
will affect the strength of the gravitational-wave driven instability. 
To do this we need to estimate the rate at which these
modes radiate gravitational waves, and also assess
the strength of various dissipation mechanisms
(like viscosity) that tend to damp an unstable mode. 
This is obviously an important issue and we plan to address
it once our ongoing work on fully relativistic 
hybrid modes of isentropic stars is completed.
At that point it will also be appropriate
to obtain and discuss results for different realistic equations
of state.


\acknowledgements

We wish to thank Kostas Kokkotas, Lee Lindblom, Sharon Morsink 
and Nick Stergioulas for helpful discussions.
This work was supported in part by NSF grants PHY95-07740 
and PHY95-14240 and by the Eberly research funds of Penn State.
NA is funded by PPARC grant PPA/G/1998/00606 in the UK.


\appendix

\section{Equations describing stationary perturbations of spherical stars}

We have derived the various equations governing stationary
perturbations of a spherical star
using the Maple tensor package by substituting expressions 
(\ref{GR:del_ep})-(\ref{GR:h_ax}) into Eqs.~(\ref{pert_Gmunu}) and 
(\ref{pert_Tmunu}) (making liberal use of the equilibrium equations
(\ref{GR_sph:tov}) through (\ref{GR_sph:mass_def}) to simplify the 
expressions). The resulting equations are listed in the 
three distinct cases $l\geq 2$, $l=1$ and $l=0$ below.

\subsection{The case $l\geq 2$}

The non-vanishing components of the perturbed Einstein equation for 
$l\geq 2$ are as follows.  We will use Eq. (\ref{thth-_l2}) below, to 
replace $H_2$ by $H_0$.  From $\delta G_t^{\ t}=8\pi\delta T_t^{\ t}$ 
we have (using primes to denote derivatives with respect to $r$) 
\bea
0 &=& e^{-2\lambda} r^2 K'' 
+ e^{-2\lambda} (3-r\lambda')rK' - \left[\half l(l+1)-1\right] K \nn\\
&& \nn \\
&& - e^{-2\lambda} rH_0'
-\left[\half l(l+1)+1-8\pi r^2\ep\right]H_0
+8\pi r^2\delta\ep.
\label{tt_l2}
\eea
From $\delta G_r^{\ r}=8\pi\delta T_r^{\ r}$ we similarly have
\bea
0 &=& e^{-2\lambda}(1+r\nu')rK' - \left[\half l(l+1)-1\right] K \nn\\
&& \nn \\
&& - e^{-2\lambda}rH_0'
+\left[\half l(l+1)-1-8\pi r^2 p\right]H_0
-8\pi r^2\delta p.
\label{rr_l2}
\eea
From 
$\delta G_\theta^{\ \theta}+\delta G_\varphi^{\ \varphi}
=8\pi\left(\delta T_\theta^{\ \theta}+\delta T_\varphi^{\ \varphi}\right)$ 
we have
\bea
0 &=& e^{-2\lambda} r^2 K'' 
+ e^{-2\lambda} \left[r(\nu'-\lambda')+2\right]rK' 
-16\pi r^2\delta p \nn\\
&& \nn \\
&& - e^{-2\lambda}r^2H_0''
- e^{-2\lambda}(3r\nu'-r\lambda'+2)rH_0'
-16\pi r^2 p H_0.
\label{thth+_l2}
\eea
From 
$\delta G_\theta^{\ \theta}-\delta G_\varphi^{\ \varphi}
=8\pi\left(\delta T_\theta^{\ \theta}-\delta T_\varphi^{\ \varphi}\right)$ 
we have
\be
H_2 = H_0.
\label{thth-_l2}
\ee
From $\delta G_r^{\ \theta}=8\pi\delta T_r^{\ \theta}$ we have
\be
K' = e^{-2\nu} \left[e^{2\nu} H_0 \right]'.
\label{rth_l2}
\ee
From $\delta G_t^{\ r}=8\pi\delta T_t^{\ r}$ we have
\be
0 = H_1 + \frac{16\pi(\epsilon+p)}{l(l+1)} e^{2\lambda} r W.
\label{tr_l2}
\ee
From $\delta G_t^{\ \theta}=8\pi\delta T_t^{\ \theta}$ we have
\be
0 = e^{-(\nu-\lambda)}
\left[e^{(\nu-\lambda)}  H_1\right]'
+ 16\pi(\ep+p)e^{2\lambda}V.
\label{tth_l2}
\ee
From $\delta G_t^{\ \varphi}=8\pi\delta T_t^{\ \varphi}$ we have
\be
h_0^{''} - (\nu'+\lambda') h_0' 
+ \left[ \frac{(2-l^2-l)}{r^2}e^{2\lambda} 
- \frac{2}{r}(\nu'+\lambda') - \frac{2}{r^2} \right] h_0 
= \frac{4}{r}(\nu'+\lambda') U.
\label{tph_l2}
\ee
From $\delta G_r^{\ \varphi}=8\pi\delta T_r^{\ \varphi}$ we have
\be
(l-1)(l+2) h_1 = 0.
\label{rph_l2}
\ee
Finally, from $\delta G_\theta^{\ \varphi}=8\pi\delta T_\theta^{\ \varphi}$ 
we have
\be
e^{-(\nu-\lambda)}
\left[e^{(\nu-\lambda)}  h_1\right]' = 0.
\label{thph_l2}
\ee

\subsection{The case $l=1$}

The $l=1$ case differs from $l\geq 2$ in two respects \cite{ct70}.  
Firstly, $H_2(r)\neq H_0(r)$, because the equation 
$\delta G_\theta^{\ \theta}-\delta G_\varphi^{\ \varphi}
=8\pi\left(\delta T_\theta^{\ \theta}-\delta T_\varphi^{\ \varphi}\right)$ 
vanishes identically.  Secondly, we may exploit the aforementioned gauge
freedom for this case to eliminate the metric functions $K(r)$ and $h_1(r)$. 
(We note that Eq.~(\ref{rph_l2}) implies $h_1(r)=0$ for $l\geq 2$ anyway.)
With these two differences taken into account the non-vanishing components 
of the perturbed Einstein equation for $l=1$ are as follows. From 
$\delta G_t^{\ t}=8\pi\delta T_t^{\ t}$ we have
\be
0 = e^{-2\lambda} rH_2' + \left(2-8\pi r^2\ep\right)H_2
-8\pi r^2\delta\ep.
\label{tt_l1}
\ee
From $\delta G_r^{\ r}=8\pi\delta T_r^{\ r}$ we have
\be
0 = e^{-2\lambda}rH_0' - H_0 + \left(1+8\pi r^2 p\right)H_2
+8\pi r^2\delta p.
\label{rr_l1}
\ee
From 
$\delta G_\theta^{\ \theta}+\delta G_\varphi^{\ \varphi}
=8\pi\left(\delta T_\theta^{\ \theta}+\delta T_\varphi^{\ \varphi}\right)$ 
we have
\bea
0 &=& e^{-2\lambda}r^2H_0''
+ e^{-2\lambda}(2r\nu'-r\lambda'+1)rH_0' - H_0 \nn\\
&& + e^{-2\lambda}(1+r\nu')rH_2' + (1+16\pi r^2 p)H_2
+16\pi r^2\delta p
\label{thth+_l1}
\eea
From $\delta G_r^{\ \theta}=8\pi\delta T_r^{\ \theta}$ we have
\be
0 = rH_0' + (r\nu'-1)H_0 + (r\nu'+1)H_2
\label{rth_l1}
\ee
From $\delta G_t^{\ r}=8\pi\delta T_t^{\ r}$ we again have
\be
0 = H_1 + 8\pi(\epsilon+p) e^{2\lambda} r W.
\label{tr_l1}
\ee
From $\delta G_t^{\ \theta}=8\pi\delta T_t^{\ \theta}$ we again have
\be
0 = e^{-(\nu-\lambda)}
\left[e^{(\nu-\lambda)}  H_1\right]'
+ 16\pi(\ep+p)e^{2\lambda}V.
\label{tth_l1}
\ee
Finally, from $\delta G_t^{\ \varphi}=8\pi\delta T_t^{\ \varphi}$ 
we have
\be
h_0^{''} - (\nu'+\lambda') h_0' 
- \left[ \frac{2}{r}(\nu'+\lambda') + \frac{2}{r^2} \right] h_0 
= \frac{4}{r}(\nu'+\lambda') U.
\label{tph_l1}
\ee

\subsection{The case $l=0$}

The $l=0$ case differs yet again from the previous two, being the 
case of stationary, spherically symmetric perturbations of a static,
spherical equilibrium.  To maximize the similarity to the preceding
two cases we will use the same form for the perturbed metric except
that we may now exploit the gauge freedom for this case to eliminate 
the functions $K(r)$, $H_1(r)$ and $h_1(r)$.  The non-vanishing 
components of the perturbed Einstein equation for $l=0$ are as follows. 

From $\delta G_t^{\ t}=8\pi\delta T_t^{\ t}$ we have
\be
0 = e^{-2\lambda} rH_2' + \left(1-8\pi r^2\ep\right)H_2
-8\pi r^2\delta\ep.
\label{tt_l0}
\ee
From $\delta G_r^{\ r}=8\pi\delta T_r^{\ r}$ we have
\be
0 = e^{-2\lambda}rH_0' + \left(1+8\pi r^2 p\right)H_2
+8\pi r^2\delta p.
\label{rr_l0}
\ee
From 
$\delta G_\theta^{\ \theta}+\delta G_\varphi^{\ \varphi}
=8\pi\left(\delta T_\theta^{\ \theta}+\delta T_\varphi^{\ \varphi}\right)$ 
we have
\bea
0 &=& e^{-2\lambda}r^2H_0''
+ e^{-2\lambda}(2r\nu'-r\lambda'+1)rH_0' \nn\\
&& + e^{-2\lambda}(1+r\nu')rH_2' + 16\pi r^2 pH_2 +16\pi r^2\delta p
\label{thth+_l0}
\eea
Finally, from $\delta G_t^{\ r}=8\pi\delta T_t^{\ r}$ we have
\be
0 = 16\pi(\epsilon+p) W.
\label{tr_l0}
\ee

\section{Perturbation equations for slowly rotating non-isentropic stars}

The assumption of a purely axial perturbation in the limit $\Omega \to 0$ 
leads to the following equations (cf. (\ref{ordering})):
The three axial quantities follow from
\begin{eqnarray}
&& \left[ \sigma +m\Omega - { 2m\bar{\omega} \over l(l+1)}\right] 
\left\{  e^{\nu-\lambda} {d\over dr} \left[ e^{-\nu-\lambda} 
{dh_{0,l} \over dr} \right] - \left[{l(l+1) \over r^2} - {4M\over r^3}
+8\pi(p+\epsilon) \right]  h_{0,l} \right\} \nonumber \\
&&+
16\pi(p+\epsilon)(\sigma+m\Omega) h_{0,l} = 0 \ .
\end{eqnarray}
\begin{equation}
\left[\sigma+m\Omega - {2m\bar{\omega} \over l(l+1)} \right] U_l + 
(\sigma+m\Omega) h_{0,l} = 0 \ . 
\end{equation}
and
\begin{equation}
 l(l+1) \left\{ i(\sigma + m\omega) e^{-2\nu}\left[  h_{0,l}^\prime - 2  
{h_{0,l} \over r} \right] -
{ (l-1)(l+2) h_{1,l} \over
r^2} \right\}
- 2im\omega^\prime e^{-2\nu} h_{0,l} = 0 \ ,
\end{equation}
or, alternatively, equation (\ref{ax2}).

The solutions to these three equations then serve as
sources for 
three of the remaining (at order $\Omega$) 
Einstein equations that determine the polar parity 
metric perturbations:

\begin{eqnarray}
&& (l-1)l(l+1)(l+2) e^{2\nu}(H_{2,l}-H_{0,l}) Y_l^m  \nonumber \\
&&-\left\{ r^2e^{-2\lambda} \omega^\prime h_{0,l}^\prime + 
\left[ l(l+1)\omega-2 r e^{-2\lambda} \omega^\prime 
- 16 \pi r^2(p+\epsilon)\bar{\omega} \right] h_{0,l} - 
16\pi i (p+\epsilon) r^2 \bar{\omega} U_l  \right\} \nonumber \\
&& \times 
\left\{ 2(l-1)(l+2)\sin \theta \partial_\theta Y_l^m +4l(l+1)
\cos\theta Y_l^m\right\}=0
\label{new2}\end{eqnarray}

\begin{eqnarray}
&& l(l+1) e^{2\nu}
\left[ r (K_l^\prime - H_{0,l}^\prime) + 
\left( 1-r\nu^\prime \right) H_{0,l} - \left(1+r\nu^\prime \right) H_{2,l}
\right] Y_l^m \nonumber \\
&& -l(l+1) \left\{ 2r\omega  h_{0,l}^\prime 
+  [ r\omega^\prime - 
2\omega(1+r\nu^\prime) ]  h_{0,l} \right\} 
\sin\theta \partial_\theta Y_l^m \nonumber \\
&&  + 2r \omega^\prime h_{0,l} (
\sin\theta \partial_\theta Y_l^m  -l(l+1)\cos \theta Y_l^m) = 0  
\label{min2}\end{eqnarray}
and
\begin{eqnarray}
&& l(l+1) e^{2\nu} 
\left[ 2\nu^\prime e^{-2\lambda} \left( K_l^\prime - 
{H_{2,l} \over r^2} \right) -
{(l-1)(l+2) \over r^2} (K_l-H_{0,l}) + 
\left( {6M \over r^3} - 8\pi \epsilon \right) H_{0,l}
\right] Y_l^m \nonumber \\
&&-l(l+1) \left[ \omega^\prime e^{-2\lambda} h_{0,l}^\prime
+ \left( 16\pi ( p\bar{\omega} +\epsilon \Omega) - \omega
{(l-1)(l+2)r + 6M\over r^3} \right) h_{0,l}
\right] \sin\theta \partial_\theta Y_l^m  \nonumber \\
&&
- \left( {4 \over r} \omega^\prime e^{-2\lambda} h_{0,l} 
- 32\pi i (p+\epsilon)\bar{\omega} U_l \right)
(\sin\theta \partial_\theta Y_l^m  -l(l+1)\cos \theta Y_l^m) \nonumber \\
&&+{4l^2(l+1)^2 \over r^2} \omega h_{0,l} \cos \theta Y_l^m = 0
\label{final}\end{eqnarray}

These equations determine the polar metric perturbations
$K_l$, $H_{0,l}$ and $H_{2,l}$ once $h_{0,l}$ and $U_l$ are known 
for all $l$. Finally, the last two Einstein equations lead to the 
following equations for $\delta p_l$ and $\delta \ep_l$
(recall that for non-isentropic stars the equation of state
does not link these two quantities until at order $\Omega^2$);
\begin{eqnarray}
&& \left[ H_{0,l}^{\prime\prime} -2\nu^\prime K_l^\prime + \left( {2\over r}+
 2\nu^\prime - \lambda^\prime \right) H_{0,l}^\prime +\nu^\prime H_{2,l}^\prime
 \right. \nonumber \\
&& \left. 
-{l(l+1) \over r^2} e^{2\lambda} H_{0,l} + 8\pi (3p+\epsilon)e^{2\lambda} H_{2,l} + 8\pi e^{2\lambda} (3\delta p_l + \delta \epsilon_l)  \right] e^{2\nu}
Y_l^m \nonumber \\
&& 
+ \left[ \left( {4\omega \over r} + 2\omega^\prime - 2\omega \nu^\prime \right) h_{0,l}^\prime + \left( 4\omega (\nu^\prime)^2 - 2\omega^\prime \nu^\prime - 
{8\omega \over r} \nu^\prime \right)  h_{0,l} + 32\pi i(p+\epsilon) \Omega 
e^{2\lambda} U_l  \right] \sin \theta \partial_\theta Y_l^m \nonumber \\
&& - { 4l(l+1) \over r^2}
\omega e^{2\lambda} h_{0,l} \cos \theta Y_l^m=0  
\label{tt}\end{eqnarray}

\begin{eqnarray}
&& 
\left[ K_l^{\prime\prime} + \left( {2\over r} - 4\pi (p+\epsilon) 
re^{2\lambda} \right) 
K_l^\prime + {1 \over r} (H_{0,l}^\prime -H_{2,l}^\prime) \right. \nonumber \\
&& \left.  - 
{l(l+1) \over 2 r^2} e^{2\lambda} (H_{0,l} +  H_{2,l}) +
 8\pi (p+\epsilon) e^{2\lambda} H_{2,l} + 8\pi e^{2\lambda}
(\delta p_l + \delta \epsilon_l) \right] Y_l^m \nonumber \\
&& 
+ \left[ 
{2 \over r} \omega e^{-2\nu} h_{0,l}^\prime + 
\left( {2\over r} (\omega^\prime  -
 2\omega \nu^\prime) +
16\pi e^{2\lambda} (p+\epsilon) \bar{\omega} - {l(l+1) \over r^2} 
\omega e^{2\lambda} \right) e^{-2\nu} h_{0,l} \right. \nonumber \\
&& \left. +16\pi i (p+\epsilon) \bar{\omega} e^{2\lambda-2\nu} U_l 
\right] \sin \theta \partial_\theta Y_l^m \nonumber \\
&&- {2l(l+1) \over r^2} \omega 
e^{2\lambda-2\nu} h_{0,l} \cos \theta Y_l^m = 0
\label{rr}\end{eqnarray}


\section{Proof of Theorem \lowercase{\ref{thm2}}}

\subsection{Axial-led hybrids with $m>0$.}

Let $l$ be the smallest value of $l'$ for which $U_{l'}\neq 0$ in 
the spherical harmonic expansion (\ref{xi_exp}) of the displacement
vector $\xi^\alpha$, or for which $h_{l'}\equiv h_{0,l'}\neq 0$ in 
the spherical harmonic expansion (\ref{h_components}) of the metric
perturbation $h_{\alpha\beta}$.  The axial parity of 
$(\xi^\alpha, h_{\alpha\beta})$, $(-1)^{l+1}$, and the vanishing of 
$Y_l^m$ for $l<m$ implies $l\geq m$.  That the mode is axial-led means 
$W_{l'}=0$, $V_{l'}=0$ and $H_{1,l'}=0$ for $l'\leq l$.  
We show by contradiction that $l=m$.

Suppose $l\geq m+1$.  From Eq. (\ref{om_th_ph}),
$\int \Delta\om_{\theta\varphi} Y_l^{\ast m} d\Omega = 0$, 
we have
\be
0 = \ba[t]{l}
l(l+1)\kappa\Omega(h_l+U_l)-2m{\bar\omega}U_l
\\ \\
-l Q_{l+1} \left[\frac{e^{2\nu}}{r}\partial_r
\left(r^2{\bar\omega}e^{-2\nu}\right)W_{l+1}
+2(l+2){\bar\omega}V_{l+1} \right],
\ea
\ee
and from Eq. (\ref{om_ph_r}) with $l\rightarrow l-1$,
$\int \Delta\om_{\varphi r} Y_{l-1}^{\ast m} d\Omega = 0$, 
we have
\be
0 = \ba[t]{l}
- \biggl\{
(l+1)\kappa\Omega\partial_r\left[e^{-2\nu}(h_l+U_l)\right]
+2m\partial_r\left({\bar\omega}e^{-2\nu}U_l\right)
+\frac{m(l+1)}{r^2}\partial_r\left(r^2{\bar\omega}e^{-2\nu}\right)U_l
\biggr\}
 \\ \\
+ Q_{l+1} \biggl[
\partial_r\left[\frac{1}{r}\partial_r
\left(r^2{\bar\omega}e^{-2\nu}\right)W_{l+1}\right]
+2(l+2)\partial_r\left({\bar\omega}e^{-2\nu}V_{l+1}\right)
\biggr]
\ea
\ee
Together these give,
\bea
0 &=& 2\partial_r\left({\bar\omega}e^{-2\nu}U_l\right)
+\frac{l}{r^2}\partial_r\left(r^2{\bar\omega}e^{-2\nu}\right)U_l
\nn \\ \\
  &=& 2\left(r^2\bom e^{-2\nu}\right)^{-\frac{l}{2}}
\partial_r\left[r^l\left(\bom e^{-2\nu}\right)^{\half(l+2)}U_l
\right]
\eea
or,
\be
U_l = K \left(\bom e^{-2\nu}\right)^{-\half(l+2)}r^{-l}
\ee
(for some constant $K$) which is singular as $r\rightarrow 0$.

\subsection{Axial-led hybrids with $m=0$.}

Let $m=0$ and let $l$ be the smallest value of $l'$ for which 
$U_{l'}\neq 0$ in the spherical harmonic expansion (\ref{xi_exp}) of 
the displacement vector $\xi^\alpha$, or for which 
$h_{l'}\equiv h_{0,l'}\neq 0$ in the spherical harmonic expansion 
(\ref{h_components}) of the metric perturbation $h_{\alpha\beta}$.  
Since $\nabla_a Y_0^0=0$, the mode vanishes unless $l\geq 1$. 
That the mode is axial-led means $W_{l'}=0$, $V_{l'}=0$ and 
$H_{1,l'}=0$ for $l'\leq l$.  We show by contradiction that $l=1$.

Suppose $l\geq 2$.  Then Eq. (\ref{om_r_th}) with $l\rightarrow l-2$,
$\int \Delta\om_{r\theta} Y_{l-2}^{\ast 0} d\Omega = 0$, 
becomes
\bea
0 &=& 2\partial_r\left({\bar\omega}e^{-2\nu}U_l\right)
+\frac{l}{r^2}\partial_r\left(r^2{\bar\omega}e^{-2\nu}\right)U_l
\nn \\ \\
  &=& 2\left(r^2\bom e^{-2\nu}\right)^{-\frac{l}{2}}
\partial_r\left[r^l\left(\bom e^{-2\nu}\right)^{\half(l+2)}U_l
\right]
\eea
or,
\be
U_l = K \left(\bom e^{-2\nu}\right)^{-\half(l+2)}r^{-l}
\ee
(for some constant $K$) which is singular as $r\rightarrow 0$.

\subsection{Polar-led hybrids with $m\geq 0$.}

Let $l$ be the smallest value of $l'$ for which $W_{l'}\neq 0$ or 
$V_{l'}\neq 0$ in the spherical harmonic expansion (\ref{xi_exp}) 
of the displacement vector $\xi^\alpha$, or for which $H_{1,l'}\neq 0$ 
in the spherical harmonic expansion (\ref{h_components}) of the metric
perturbation $h_{\alpha\beta}$.  The polar parity of 
$(\xi^\alpha, h_{\alpha\beta})$, $(-1)^l$, and the vanishing of $Y_l^m$ 
for $l<m$ implies $l\geq m$.  That the mode is polar-led means 
$U_{l'}=0$ and $h_{l'}=0$ for $l'\leq l$.  
We show by contradiction that $l=m$ when $m>0$ and that $l=1$ when $m=0$.

Suppose $l\geq m+1$.  From Eq. (\ref{om_th_ph}) with 
$l\rightarrow l-1$,
$\int \Delta\om_{\theta\varphi} Y_{l-1}^{\ast m} d\Omega = 0$, 
we have
\be
0 = (l-1) Q_l \left[\frac{e^{2\nu}}{r}\partial_r
\left(r^2{\bar\omega}e^{-2\nu}\right)W_l
+2(l+1){\bar\omega}V_l \right].
\label{polar_proof}
\ee

Substituting for $V_l$ using Eq. (\ref{GR:sph_V}), we find
\bea
0 &=& \frac{l}{r}\partial_r\left(r^2{\bar\omega}e^{-2\nu}\right)W_l
+ 2\bom e^{-2\nu}\frac{e^{-(\nu+\lambda)}}{(\ep+p)}
\partial_r\left[(\ep+p)e^{(\nu+\lambda)}rW_l\right]
\nn \\ \\
 &=&2\left(r^2{\bar\omega}e^{-2\nu}\right)^{-\half(l-2)}
\frac{e^{-(\nu+\lambda)}}{r^2(\ep+p)}
\partial_r\left[\left(r^2{\bar\omega}e^{-2\nu}\right)^{\frac{l}{2}}
(\ep+p)e^{(\nu+\lambda)}rW_l\right]
\eea
with solution,
\be
W_l= K \left(\bom e^{-2\nu}\right)^{-\frac{l}{2}}
\frac{e^{-(\nu+\lambda)}}{(\ep+p)}r^{-(l+1)}
\ee
(for some constant $K$) which is singular as $r\rightarrow 0$.

When $m=0$ this argument fails to establish that $l$ cannot be equal to 
1, because Eq. (\ref{polar_proof}) is trivially satisfied for $l=1$ as a 
result of the overall $l-1$ factor.  Instead, the argument proves that 
$l$ cannot be greater than 1 in this case and therefore that $l=1$.



\newpage

\newpage

\newpage


\begin{table*}
\caption{Relativistic r-mode and hybrid mode frequencies of 
uniform density stars.  We list the numerically determined values 
of the post-Newtonian r-mode frequency correction, 
$\kappa_{1_{\mbox{\scriptsize r-mode}}}$, for the non-isentropic star and 
compare the corresponding eigenvalues $\alpha_{pN}$ to ones obtained 
for fully relativistic uniform density stars, $\alpha$, for a star 
of compactness $2M_0/R=0.4$. In this case the value of the frame dragging 
at the surface of the star leads to $\bar{\omega}(R)/\Omega = 0.84424$ 
and we can see that the eigenvalues approach this value as $l$ increases. 
It is also interesting to compare our post-Newtonian eigenvalues to the 
result we deduce for the hybrid $l=m$ modes of isentropic stars, 
$\kappa_{1_{\mbox{\scriptsize hybrid}}} =-4(m-1)(2m+11)/5(2m+1)(2m+5)$. 
The two results typically do not differ by more than a few percent.
This is important since the two modes (for $l=m$) correspond to the 
relativistic analogue (for non-isentropic and isentropic stars, 
respectively) of the same Newtonian r-mode.}
\begin{tabular}[t]{ccccc}
$l$ & $\kappa_{1_{\mbox{\scriptsize hybrid}}}$ & 
$\kappa_{1_{\mbox{\scriptsize r-mode}}}$ & $\alpha_{pN}$ & $\alpha$ \\ 
\tableline
2  & -0.2667 & -0.2629 & 0.8949 & 0.9086 \\
3  & -0.3532 & -0.3428 & 0.8629 & 0.8699 \\
4  & -0.3897 & -0.3734 & 0.8506 & 0.8561 \\
5  & -0.4073 & -0.3868 & 0.8453 & 0.8502 \\
6  & -0.4163 & -0.3931 & 0.8428 & 0.8474 \\
7  & -0.4211 & -0.3962 & 0.8415 & 0.8460 \\
8  & -0.4235 & -0.3979 & 0.8408 & 0.8453 \\
9  & -0.4247 & -0.3988 & 0.8405 & 0.8448 \\
10 & -0.4251 & -0.3993 & 0.8403 & 0.8446 \\
\end{tabular}
\label{tab1}
\end{table*}
\end{document}